%% file: PORB.tex
\documentclass{aa}
\usepackage{graphicx}
\usepackage{natbib}
\usepackage{lscape}
\usepackage{longtable}
\usepackage{verbatim}
\usepackage{subfigure}

\newcommand{\gppr}{\stackrel{>}{\scriptstyle \sim}}
\newcommand{\gappr}{\raisebox{-0.4ex}{$\gppr$}}
\newcommand{\lppr}{\stackrel{<}{\scriptstyle \sim}}
\newcommand{\lappr}{\raisebox{-0.4ex}{$\lppr$}}

\newcommand{\Porb}{\mbox{$P_\mathrm{orb}$}}

\newcommand{\Mwd}{\mbox{$M_\mathrm{wd}$}}
\newcommand{\Msec}{\mbox{$M_\mathrm{sec}$}}
\newcommand{\Ksec}{\mbox{$K_\mathrm{sec}$}}

\newcommand{\Rsec}{\mbox{$R_\mathrm{sec}$}}

\newcommand{\Ha}{\mbox{${\mathrm H\alpha}$}}
\newcommand{\Na}{\mbox{${\mathrm {Na\,I}}$}}
\newcommand{\Msun}{\mbox{$\mathrm{M}_{\odot}$}}
\newcommand{\Rsun}{\mbox{$R_{\odot}$}}
\newcommand{\Teff}{\mbox{$T_\mathrm{eff}$}}

\newcommand{\logg}{\mbox{$log\mathrm{(g)}$}}
\newcommand{\dsec}{\mbox{$d_{\mathrm{2}}$}}
\newcommand{\dwd}{\mbox{$d_{\mathrm{wd}}$}}
\newcommand{\kms}{\mbox{$\mathrm{km\,s}^{-1}$}}

\begin{document}

\title{Post common envelope binaries from SDSS. XII:\\ The orbital
  period distribution}
\authorrunning{Nebot G\'omez-Mor\'an et al.}
\titlerunning{The orbital period distribution of SDSS PCEBs}
\author{
A.~Nebot G\'omez-Mor\'an\inst{1,2},
B.~T. G\"ansicke\inst{3},
M.~R. Schreiber\inst{4}, 
A.~Rebassa-Mansergas\inst{4},
A.~D. Schwope\inst{2},
J.~Southworth\inst{5},
A.~Aungwerojwit\inst{6,7},
M.~Bothe\inst{2},
P.~J. Davis\inst{8},
U.~Kolb\inst{9},  
M.~M\"uller\inst{2},
C.~Papadaki\inst{10}, 
S.~Pyrzas\inst{3},
A.~Rabitz\inst{2},
P.~Rodr\'iguez-Gil\inst{11,12,13},
L.~Schmidtobreick\inst{14},
R.~Schwarz\inst{2},
C.~Tappert\inst{4},
O.~Toloza\inst{4},
J.~Vogel\inst{2},
M.~Zorotovic\inst{4,15}
}
\institute{
Universit\'e de Strasbourg, CNRS, UMR7550, Observatoire Astronomique de Strasbourg, 11 Rue de l'Universit\'e, F-67000 Strasbourg, France\\
\email{ada.nebot@astro.unistra.fr}
\and
Leibniz-Institut f\"ur Astrophysik Potsdam, 
An der Sternwarte 16, D-14482 Potsdam, Germany
\and
Department of Physics, University of Warwick, Coventry CV4 7AL, UK
\and
Departamento de F\'isica y Astronom\'ia, Facultad de Ciencias, Universidad 
de Valpara\'iso, Valpara\'iso, Chile 
\and
Astrophysics Group, Keele University, Staffordshire, ST5 5BG, UK
\and
Department of Physics, Faculty of Science, Naresuan University, Phitsanulok 65000, Thailand
\and
ThEP Centre, CHE, 328 Si Ayutthaya Road, Bangkok, 10400, Thailand
\and
Institut d'Astronomie et d'Astrophysique, Universit\'e Libre de Bruxelles, CP226, Boulevard du Triomphe, B-1050, Belgium
\and
Open University, dept. Physics \& Astronomy, Milton Keynes MK7 6BJ, UK
\and
Institute of Astronomy \& Astrophysics, National Observatory of
 Athens, 15236 Athens, Greece
\and 
Instituto de Astrof\'isica de Canarias, V\'ia L\'actea, s/n, La Laguna, E-38205, Tenerife,  Spain
\and
Departamento de Astrof\'isica, Universidad de La Laguna, Avda. Astrof\'isico Fco. S\'anchez, s\/n, La Laguna, E-38206, Tenerife, Spain
\and
Isaac Newton Group of Telescopes, Apartado de correos 321, S/C de la Palma, E-38700, Canary Islands, Spain
\and
European Southern Observatory, Alonso de C\'ordova 3107, Vitacura, Santiago, Chile
\and
Departamento de Astronom\'ia y Astrof\'isica, Pontificia Universidad Cat\'olica, Vicu\~na Mackenna 4860, 782-0436 Macul, Chile
}
%\offprints{A.Nebot G\'omez-Mor\'an}

\date{Received / Accepted }

\abstract
{The complexity of the common envelope phase and of magnetic stellar
  wind braking currently limits our understanding of close binary
  evolution. Because of their intrinsically simple structure,
  observational population studies of white dwarf plus main sequence
  (WDMS) binaries hold the potential to test theoretical models and
  constrain their parameters.}
{The Sloan Digital Sky Survey (SDSS) has provided a large and
  homogeneously selected sample of WDMS binaries, which we are
  characterising in terms of orbital and stellar parameters.}
{We have obtained radial velocity information for 385 WDMS binaries
  from follow-up spectroscopy, and for an additional 861 systems from the
  SDSS sub-spectra. Radial velocity variations identify 191 of these
  WDMS binaries as post common envelope binaries
  (PCEBs). Orbital periods of 58 PCEBs were subsequently measured,
  predominantly from time-resolved spectroscopy, \mbox{bringing} the total
  number of SDSS PCEBs with orbital parameters to 79. Observational
  biases inherent to this PCEB sample were evaluated through extensive
  Monte Carlo simulations.}
{We find that $21-24$\% of all SDSS WDMS binaries have undergone
  common envelope evolution, which is in good agreement with published
  binary population models and high-resolution \textit{HST} imaging of
  WDMS binaries unresolved from the ground. The bias corrected orbital 
  period distribution of PCEBs ranges from 1.9\,h to 4.3\,d and follows 
  approximately a normal
  distribution in $\log(\Porb)$, peaking at $\sim10.3$\,h. There is no
  observational evidence for a significant population of PCEBs with
  periods in the range of days to weeks. }
 {The large and homogeneous sample of SDSS WDMS binaries provides
   the means to test fundamental predictions of binary population
   models, and hence to observationally constrain  the evolution of all 
   close compact binaries. } 

\maketitle

\section{Introduction}
It is well known that all types of close compact binaries,
including X-ray binaries, cataclysmic variables, and double
degenerates, form through common envelope evolution
\citep{paczynski76-1}. This phase is required to explain
the observed small binary separations in compact binaries.  
The main concept is that, for a suitable range of orbital separations, 
the massive star (i.e. the primary) will fill its Roche lobe as it evolves up 
the giant (or asymptotic giant) branch, and will then in most cases become 
susceptible to dynamically unstable mass transfer.
The transferred material cannot cool as fast as it is being transferred 
and the core of the primary star and the companion star (also referred to as 
secondary star) are both engulfed by
the envelope of the former.  The envelope is not co-rotating with the binary, creating drag forces which transport
orbital angular momentum and energy from the binary to the common
envelope, leading to a shrinkage of the orbit \citep{ricker+taam08-1}.
Eventually, the common envelope is ejected, leaving behind a post common 
envelope binary (PCEB).

Although the basic concept of the common envelope phase has been
outlined 30 years ago, it is still the most poorly understood phase of
compact binary evolution. While we have learnt that this phase must
be very short ($\le 10^3$ yrs) \citep{hjellming+taam91-1,
  webbink08-1}, current hydrodynamical simulations of common envelope
evolution are still unable to unambiguously link the initial
parameters of the binary with the outcome of the common envelope, and
numerically too expensive to be run on large ranges of initial binary
parameters \citep{ricker+taam08-1}. Therefore theoretical binary 
population synthesis studies (BPS)
usually adopt a simple parametrisation: a certain fraction
($\alpha_\mathrm{CE}$) of the binary's binding energy which is
released in the spiralling-in process is used to unbind the common
envelope \citep{webbink84-1}. However, \citet{nelemansetal00-1} and 
\citet{nelemans+tout05-1} find difficulties even with this scaled energy 
relation when trying to reproduce the evolution of double white dwarf binaries, 
and propose a prescription based on angular momentum conservation. 
A fundamental consequence of this approach is that
clear observational constraints on these parameters are necessary to
make any meaningful predictions on the Galactic population of compact
binaries. 

PCEBs\footnote{For the remainder of the paper, we will use
  \textit{PCEB} as a synonym for a short-period WDMS binary that
  underwent a common envelope phase.} composed of a white dwarf and a main-sequence
(WDMS) companion represent the most
promising population to derive such observational constraints, as they
are the intrinsically most abundant population of PCEBs and hence
easily accessible to intense studies using the current suite of
2--8\,m teles\-copes, they underwent only one common envelope phase, are
not altered by ongoing accretion, and their stellar components are
structurally simple. In general, WDMS binaries comprise two
fundamentally different types of systems: wide binaries that evolved
like single stars, i.e.  without interaction, and PCEBs.  The major
shortcoming of using PCEBs for studying compact binary evolution has been 
their small number. The comprehensive study of
\citet{schreiber+gaensicke03-1} contained only 30 systems with
accurate binary parameters, and highlighted that the sample of PCEBs
known at that time was not only small but also heavily biased towards
hot white dwarfs and late se\-con\-da\-ry star spectral types. This
bias was identified by \citet{schreiber+gaensicke03-1} as a natural
consequence of the fact that most of these PCEBs were discovered in
blue-excess surveys.

Over the past few years, the Sloan Digital Sky Survey (SDSS,
\citealt{yorketal00-1, abazajianetal09-1, yannyetal09-1}) has totally
changed this situation. Thanks to its broad $ugriz$ colour space, and
massive spectroscopic follow-up, SDSS has identified over 1900 WDMS
binaries \citep{smolcicetal04-1, silvestrietal06-1, silvestrietal07-1,
  schreiberetal07-1, helleretal09-1, rebassa-mansergasetal07-1,
  rebassa-mansergasetal10-1}. We have ini\-tiated an intensive follow-up
study of the SDSS WDMS binaries, with the ultimate aim to
substantially improve our understanding of compact binary
evolution. First results on individual systems and small sub-sets of
this sample have been published  \citep{schreiberetal08-1,
  rebassa-mansergasetal08-1, nebotgomez-moranetal09-1, pyrzasetal09-1,
  schwopeetal09-1}, as well as several studies that analyse the total
sample in the context of close binary evolution
\citep{schreiberetal10-1, zorotovicetal10-1,
  rebassa-mansergasetal11-1}. Here we present one of the fundamental
results of this project: the orbital period distribution of 79 SDSS
WDMS binaries, which we discuss in the context of current theories of
compact binary evolution.

\section{Overall strategy and input sample}
\label{sec:sample}

Our aim of characterising a large and homogeneous sample of PCEBs
requires a multi-stage approach. The first step is to identify all
WDMS binaries within the SDSS spectroscopic data release. The latest
catalogues of WDMS binaries from SDSS list 1903 WDMS systems, $\sim1600$
from \citet{rebassa-mansergasetal10-1} based on the SDSS Data Release~(DR)~6,
plus another $\sim300$ that have 
been discovered by us \citep{schreiberetal07-1} as part of the Sloan Extension 
for Galactic Understanding and Exploration (SEGUE, \citealt{yannyetal09-1}). 
These cata\-logues include white dwarf temperatures and masses,
as well as the spectral types of the companion stars.  
In brief, stellar parameters are derived from fitting the SDSS spectrum to template spectra made of white dwarf plus dM templates. The spectral type of the secondary star is derived from the best fit and the mass is calculated using a spectral type-mass relation \citep[see][for details]{rebassa-mansergasetal07-1}. The mass and the temperature of the white dwarf are derived from fitting both lines and continuum (after the contribution from the companion has been subtracted) to stellar models from \cite{koesteretal05-1}. 

The second step
is to identify the PCEBs among the entire sample of WDMS binaries. The
comprehensive population synthesis by \citet{willems+kolb04-1}
suggested that $\sim75\%$ of all Galactic WDMS binaries are wide
systems, and that the vast majority of the \mbox{remaining} $\lappr\,25\%$
underwent a common envelope phase, hence several hundred PCEBs are to
be expected among the SDSS WDMS sample.  A small fraction of WDMS
binaries appear as blended white dwarf plus M-dwarf pairs in the SDSS
imaging, and were flagged as wide systems by
\citet{rebassa-mansergasetal10-1}. However, even those WDMS binaries
that are genuinely unresolved on the SDSS images can still have
orbital separations of many tens of astronomical units. The key to
identify the close WDMS binaries is hence to carry out a radial
velocity survey of the spatially unresolved SDSS WDMS binaries,
obtaining at least two spectra, separated by at least one night, of as
many WDMS binaries as possible. In a final step, intense follow-up
spectroscopy of the PCEBs, i.e.  WDMS binaries found to be radial
velocity variable, is needed to measure their orbital periods.

\section{Observations, reduction, and radial velocity measurements}
\label{sec:observations}

We obtained spectroscopic follow-up observations for 385 systems
(Sect.\ref{sec:spec-follow}), and used exclusively SDSS spectroscopic
data for another 861 systems, bringing the number of SDSS WDMS
binaries with at least two reliable radial velocity measurements to
1246.  In addition, we obtained photometric time-series for seven systems
(Sect.\ref{sec:phot-follow}).  A log of the observations is given in
the online version of this article (Table~\ref{t:log}), listing the name
of the object, the \emph{ugriz} magnitudes, the telescopes, the number
of spectra taken and the duration of the observations.

\onltab{1}{
\input{table1.tex}
}

\subsection{Spectroscopy}
\label{sec:spec-follow}
Throughout the period January 2006 to July 2010, we have used the
3.5\,m telescope at Calar Alto (CA), the 3.5 New Technology Telescope
(NTT), the 4.2\,m William Herschel Telescope (WHT), the two 6.5\,m
Magellan telescopes (Baade/Clay), the two 8\,m Gemini telescopes
(GN/GS), and the 8\,m Very Large Telescope (VLT) to obtain
spectroscopy of 385 SDSS WDMS binaries. Details on the instrumentation
and setup are given in Table~\ref{t:res}.

The data obtained at Calar Alto were reduced within MIDAS\footnote{http://www.eso.org/sci/software/esomidas} and the
spectra were extracted using the optimal algorithm
\citep{horne86-1}. The data from all other telescopes was reduced
using the STARLINK\footnote{starlink.jach.hawaii.edu/starlink} 
packages FIGARO and KAPPA, spectra were optimally
extracted using PAMELA \citep{marsh89-1} and wavelength-calibrated
within MOLLY\footnote{www.warwick.ac.uk/go/trmarsh/software}. For all
the spectra, the wavelength calibration was checked, and if necessary
corrected, for telescope/instrument flexure using the night sky
lines. The spectra were flux calibrated and corrected for telluric
absorption using observations of the stars BD\,+28$^\circ$4211,
BD\,+33\,+$^\circ$2642, BD\,+25$^\circ$4655, Feige\,66, Feige\,110,
GD\,108, and LTT\,3218. For additional details on the reduction and
calibration of the data, see \cite{schreiberetal08-1,
  rebassa-mansergasetal08-1, nebotgomez-moranetal09-1}.

In addition to our own follow-up spectroscopy, we made use of the
spectroscopic data from SDSS. Every SDSS spectrum is the average of
(typically) three 15\,min exposures. These \textit{sub-spectra} are
publically available since DR6 \citep{adelman-mccarthyetal08-1}. As shown by
\cite{rebassa-mansergasetal08-1} and \cite{schwopeetal09-1} it is
possible to use many of these sub-spectra to measure radial velocities
and identify short-period WDMS binaries. We retrieved the subspectra
for all 1903 systems in our input catalogue. Discarding spectra of
insufficient signal-to-noise ratio, we were able to measure at least
one \mbox{radial} velocity for 1147 WDMS binaries, and we measured a total of
5171 radial velocities using the SDSS data.  
The wavelength calibration of the SDSS spectra can be accurate to 
$\simeq2$\,\kms\, in dark nights, while in bright nights there can be shifts up to 20\,\kms. 
A systematic error of 7 \kms\, has been reported by \citet{stoughtonetal02-1} and \citet{yannyetal09-1}. 
Although the SEGUE spectroscopic data has been corrected for that offset, 
this shift has only been applied to the combined spectra and not the 
individual subspectra.
We take this into account by adding in quadrature 10\,\kms\, to the statistical errors 
on the radial velocities measured from Gaussian fits (see Sect.~\ref{sec:rv_measur}).

\input{table2.tex}

\subsection{Radial velocity measurements}
\label{sec:rv_measur}
We measured the radial velocities of the companion stars by \mbox{fitting} a
second-order polynomial plus two Gaussians to the observed
$8183/8194$\,\AA\, Na\,I absorption doublet. The separation of the two
Gaussians was fixed to the laboratory value of the Na\,I doublet.  We
also measured the radial velocities from the \Ha\, \mbox{emission} line
(where present) by fitting a second-order polynomial to the underlying
continuum, plus a Gaussian for the \mbox{emission} line 
\citep[see][for further details on the method]{rebassa-mansergasetal07-1,rebassa-mansergasetal08-1}. 
Radial velocities,
errors, time of the observations and the telescope used for the 
identification of the system are available via the CDS; an excerpt is 
presented in Table\,\ref{t:rvs_iden}.

\subsection{Photometric follow-up observations} 
\label{sec:phot-follow}

We carried out time-series photometry for 7 WDMS binaries to probe for
orbital variability (Table\,\ref{t:log}) where the available
spectroscopy suggested a short orbital period. These data were
obtained between September 2006 and August 2009 using the 2.5\,m
DuPont telescope, the 2.2\,m telescope at Calar Alto, and the 0.8\,m
IAC\,80 telescope. All observations were carried out with red filters
(Bessel $R$, Sloan $r$ or $i$) to maximise the sensitivity to changes
in the brightness of the secondary star (ellipsoidal modulation and/or
reflection effect). The IAC80 data were reduced using standard packages
in IRAF\footnote{IRAF is distributed by the National Optical Astronomy
  Ob\-ser\-va\-to\-ry, which is operated by the Association of
  Universities for Research in Astronomy, Inc., under contract with
  the National Science Foundation, http://iraf.noao.edu}. The DuPont
and Calar Alto data were de-biased and flat-fielded in a standard
fashion within MIDAS, and instrumental magnitudes were derived \mbox{using}
SExtractor \citep{bertin+arnouts96-1}. For a full description of
the photometric reduction pipeline see \citet{gaensickeetal04-2}.

The results of the photometric observations are discussed in
Sect.\,\ref{sec:porb_photo} and \ref{sec:incl}, and details on
individual systems are given in the online version
(Appendix\,\ref{sec:app2}).

\section{Observational results}

\subsection{Post common envelope binary identification}
\label{sec:subspectra_ana}

Our observational program started back in 2005. The original strategy
for identification of PCEBs was to take two to three spectra for each
WDMS binary at random times, with the only constraint that they should
be separated by at least one night. Any system displaying significant radial
velocity variations was then flagged for intense follow-up
spectroscopy to measure its orbital period.

With the wealth of SDSS increasing with every data release, our
strategy has evolved. Within DR5, \cite{rebassa-mansergasetal07-1}
found that $\sim100$ WDMS had multiple SDSS spectra, and identified a
certain fraction of PCEBs among them. As mentioned in 
Sect.~\ref{sec:spec-follow} each SDSS spectrum is typically
the \mbox{average} of three individual exposures.
Occasionally, the subspectra were obtained
on different nights, and for a small number of objects more than three
subspectra are available. These subspectra were made publically
available from DR6 onwards, and we made use of them for the
identification of PCEBs. For $\sim100$ systems, the signal-to-noise
ratio of the subspectra was insufficient for an accurate radial
velocity measurement. In these cases, we combined the radial velocity
determined from the co-added SDSS spectra with our own
measurements. For those WDMS binaries that have SDSS data obtained on
different nights, we usually took only one additional spectrum to have 
an independent measurement.

\begin{figure}[t!]
\begin{center}
\includegraphics[angle=90,width=0.5\textwidth]{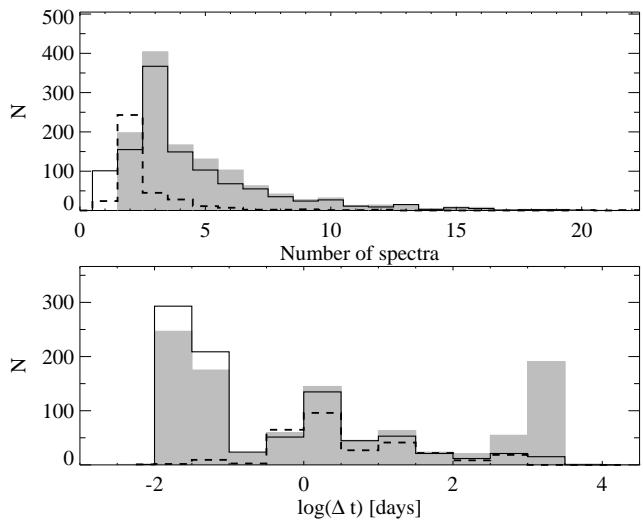}
\caption[]{Upper panel:  frequency of the number of  spectra taken per
  system from  our spectroscopic campaign (dashed), from  the SDSS (solid)
  and  total   combined  number  of  spectra   (gray).  Bottom  panel:
  distribution of  time span  between the first  and the  last spectrum
  taken   for    identification   of   close    and   wide   binaries.
  See text for a detailed explanation.}\label{g:num_spec}
\end{center}
\end{figure}

Figure\,\ref{g:num_spec} shows the distribution of the number of
spectra per system used to separate PCEBs and wide binary candidates
(upper panel), and the time elapsed between the first and last
spectrum for a given object (bottom panel), with the SDSS data 
(solid line), our follow-up data (dashed line), and the combined data (gray).  
Most of the WDMS binaries have 3 SDSS subspectra which were taken in the same
night, and are hence sensitive only to short orbital periods. A second
peak around a few days indicates SDSS repeat observations of the same
spectroscopic plate. Our own spectroscopy was typically carried out with a
separation of $\sim1-3$\,days in visitor mode observing runs, with a
long tail coming from service mode observations on the VLT and on
Gemini. We make the reader note that the combined distributions 
are not the direct sum of the other two, there are systems 
that were observed only once by us but had several SDSS subspectra, 
or the other way round. 

For the identification of radial-velocity variable stars, i.~e. PCEBs,
we analysed the radial velocities measured from the Na\,I
doublet and the H$\alpha$ emission line separately, since these two
velocities can differ by more than the statistical errors
\citep{rebassa-mansergasetal07-1}, e.g. if H$\alpha$ originates
predominantly in the inner hemisphere of the companion.  We followed
the approach of \cite{maxtedetal00-1} to evaluate the significance of
the measured radial velocity differences.  We calculated for each
system the $\chi^2$ statistic with respect to the mean radial
velo\-city, which represents the best fit for a non-variable system.
For the hypothesis of a constant radial velocity, the \mbox{probability} $Q$
of obtaining a given $\chi^2$ is high when radial velocity \mbox{variations}
are dominated by random noise, and low for intrinsically
radial-velocity variable systems.  The probability of a system showing
high radial velocity variations is hence given by $P(\chi^2)=1-Q$. We consider
strong PCEB candidates those systems showing radial velocity
variations at a $>3\sigma$ significance, or $P(\chi^2)>99.73$\%. The number
of spectra used for each system, time span covered by these spectra,
and the probabilities $P(\chi^2)$ are availa\-ble via the CDS; for an extract see 
Table~\ref{t:stats}.

\begin{figure}[t!]
\begin{center}
\includegraphics[angle=90,width=0.5\textwidth]{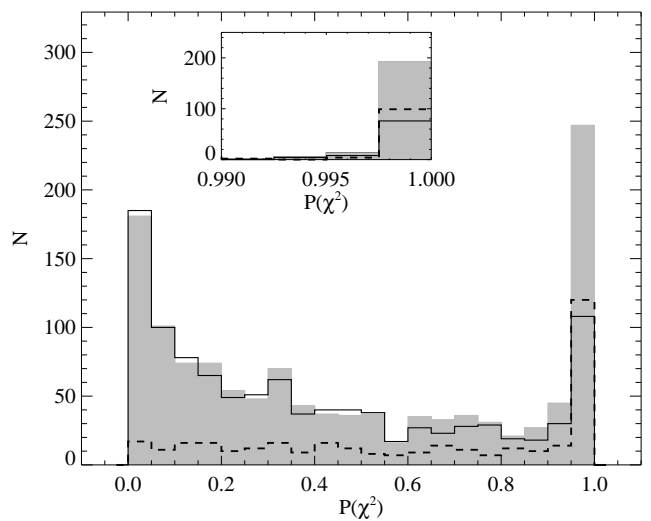}
\caption[]{Distribution of  the probability of  measuring strong
radial  velocity variations among  the WDMS binaries  in the  sample. Those
above  a  $3\sigma$ ($P(\chi^2)>99.73$) threshold   are  considered  to  be PCEBs. 
Line coding  as in Fig.~\ref{g:num_spec}. 
A zoom in the region between  0.995 and 1 is
shown in the box, note that the bin size is smaller. For a detailed 
explanation see text.}\label{g:probab}
\end{center}
\end{figure}

\input{table3.tex}
\input{table4.tex}

The distribution of $P(\chi^2)$ is shown in Fig.~\ref{g:probab}, again for
the SDSS data (solid line), our follow-up data (dashed line), and the combined
data (gray). The combined probability distribution presents a
pronounced peak at $P(\chi^2)=1$, corres\-ponding to systems exhibiting a
statistically significant radial velocity variation, i.~e.  PCEBs, and
a less pronounced one close to $P(\chi^2)=0$, corres\-ponding mainly to very
wide systems, and a small contribution of PCEBs with very low
inclinations resulting in projected radial velocities that are too low
to be picked up by the spectroscopic data. In total we found 191 PCEBs
among the 1246 SDSS WDMS binaries for which at least two radial
velocity measurements are available. 
These numbers can be used to set a
strict lower limit on the fraction of PCEBs, see
Sect.\,\ref{sec:pcebfraction} for a more detailed analysis.
For clarity, as for Fig.~\ref{g:num_spec}, we mention that the combined distribution of 
$P(\chi^2)$ is not the direct sum of the individual sets and illustrate this
with an example: SDSS\,0848+0058 
has three SDSS subspectra taken within 1.4 hours, and we took two spectra separated by 
five nights. We obtained $P(\chi^2)=0.4573$ and $P(\chi^2)=0.4014$ respectively. Combining
the data we have 5 spectra, separated by 7 years and obtain $P(\chi^2)=0.7140$.

It is clear that any radial velocity survey will be subject to
observational biases related to the spectral resolution employed, the
number of spectra obtained for each object, and the temporal spacing
of these spectra. These biases are evaluated in
Sect.\,\ref{sec:biases}. They are also responsible for the differences
between the probability distribution from the SDSS data and from our
own follow-up observations, which are discussed in
Sect.\,\ref{sec:porb_dis}.

\subsection{Orbital period measurements}
As mentioned above, 191 WDMS binaries exhibited significant radial
velocity variations and were flagged for detailed follow-up
observations. 15 orbital periods  were already published by
\cite{pyrzasetal11-1,schreiberetal08-1, rebassa-mansergasetal08-1,
  nebotgomez-moranetal09-1} and \cite{pyrzasetal09-1}. Here we report
period determinations for 58 additional systems, four of which were
measured from follow-up photometry, the rest from follow-up 
spectroscopy. 

\subsubsection{Orbital periods from photometric observations}
\label{sec:porb_photo}
Photometric variability in PCEBs can be caused by four diffe\-rent
mechanisms: ellipsoidal modulation due to the distorted shape of the
secondary star, reflection effect on the surface of the secondary star
due to irradiation from the hot primary, eclipses, and rotational 
modulation due to starspots (BY Dra phenomenon). The first three types of 
variability are, by definition, modulated on the orbital period. For most 
PCEBs, also variability due to star spots will be modulated on the orbital 
period, as the companion will by tidally synchronised. However, for wide 
binaries, star spots on the companion star may mimic variability of a  
short-period PCEB.  Four out of the
seven WDMS binaries for which we obtained time-series photometry
(Table\,\ref{t:log}) show ellipsoidal modulation, and we determined
their orbital periods by fitting sine curves to the observations.
Periodograms and phase folded light curves are presented and discussed
in more detail in Appendix~\ref{sec:app2}.

\subsubsection{Orbital periods from spectroscopic observations}
\label{sec:Porb}
We obtained further spectroscopy for 68 of the 191 PCEBs iden\-ti\-fied by
our radial velocity survey (Table\,\ref{t:log}). A pe\-riod search was
per\-for\-med by com\-pu\-ting pe\-rio\-do\-grams adop\-ting
\citeauthor{schwarzenberg-czerny96-1}'s (1996) ORT/TSA method within
MIDAS, which fits periodic orthogonal polynomials to the observations,
combined with an analysis of variance statistic. In a second step, we
fitted sine curves of the form
\begin{equation}
v_r = \gamma_{\mathrm{sec}} + K_{\mathrm{sec}} \sin\left[\frac{2\pi(t-t_\mathrm{0})}{\Porb}\right], 
\end{equation}
to the radial velocity data, with $\gamma_{\mathrm{sec}}$ the systemic
velocity of the secondary star, $t_\mathrm{0}$ the zero point defined
by the inferior conjunction of the secondary star, and
$K_{\mathrm{sec}}$ the radial velocity amplitude of the secondary
star. Initial values for $\Porb$ were selected from the highest peaks in
the periodograms. We adopt as orbital period the values of the
best-fit solution, which was always coinciding with the highest peak
in the periodogram. We could measure orbital periods for 58 of the 68
PCEBs that were spectroscopically followed up. For the remaining 10
systems the radial velocity data was insufficient to unambiguously
identify the true orbital periods. The radial velocity variations 
measured for these 10 systems range from 50 \kms\, to 490 \kms\, so a priori 
there is no reason to think that their orbital period distribution differs 
significantly from the one drawn from the rest of the population of PCEBs.

The 58 orbital periods range from $1.97$ to $104.56$ hours (0.08 to 
4.35 days) (Table~\ref{t:allparam}). The spectra of these PCEBs, and their
phase-folded radial velocity curves are shown in
Fig.~\ref{g:spectra1},~\ref{g:spectra2} and Fig.~\ref{g:rv1},~\ref{g:rv2}, respectively, 
and their
stellar and binary parameters given in Table~\ref{t:allparam}, which
also lists $K_{\mathrm{sec}}$ and $\gamma_{\mathrm{sec}}$.  The
orbital period distribution of the SDSS PCEBs is discussed in more
detail in Sect.\,\ref{sec:porb_dis}.

\onltab{5}{
\input{table5.tex}
}

\onlfig{3}{
\begin{figure*}
\begin{center}
\includegraphics[angle=0,clip=,height=\textheight]{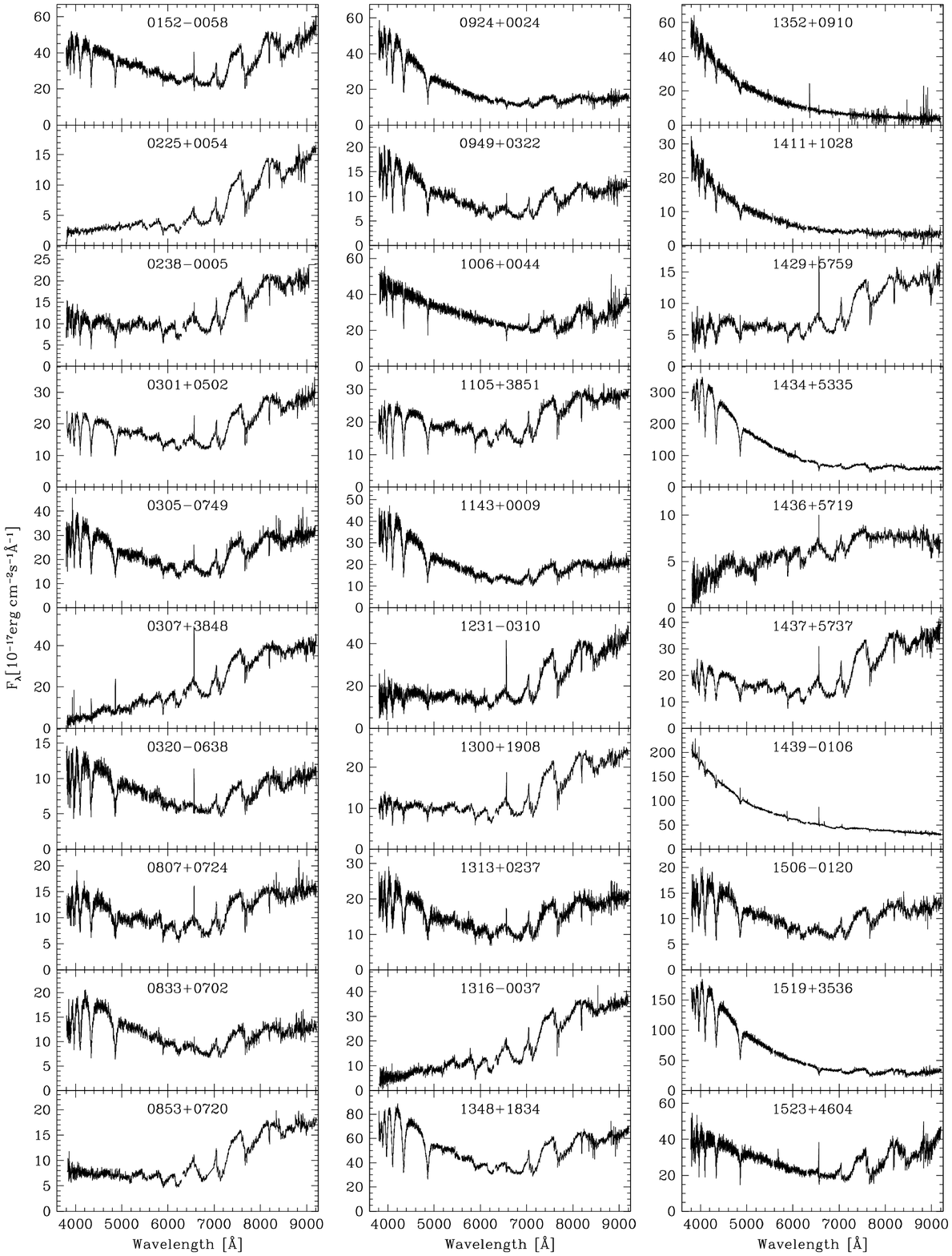}
\caption [SDSS  spectra]{SDSS spectra of  the 58 systems  with orbital
period measurement. Systems are sorted in right ascension from left to
right and from top to bottom.}\label{g:spectra1}
\end{center}
\end{figure*}
}%
\onlfig{4}{
\begin{figure*}
\begin{center}
\includegraphics[angle=0,clip=,height=\textheight]{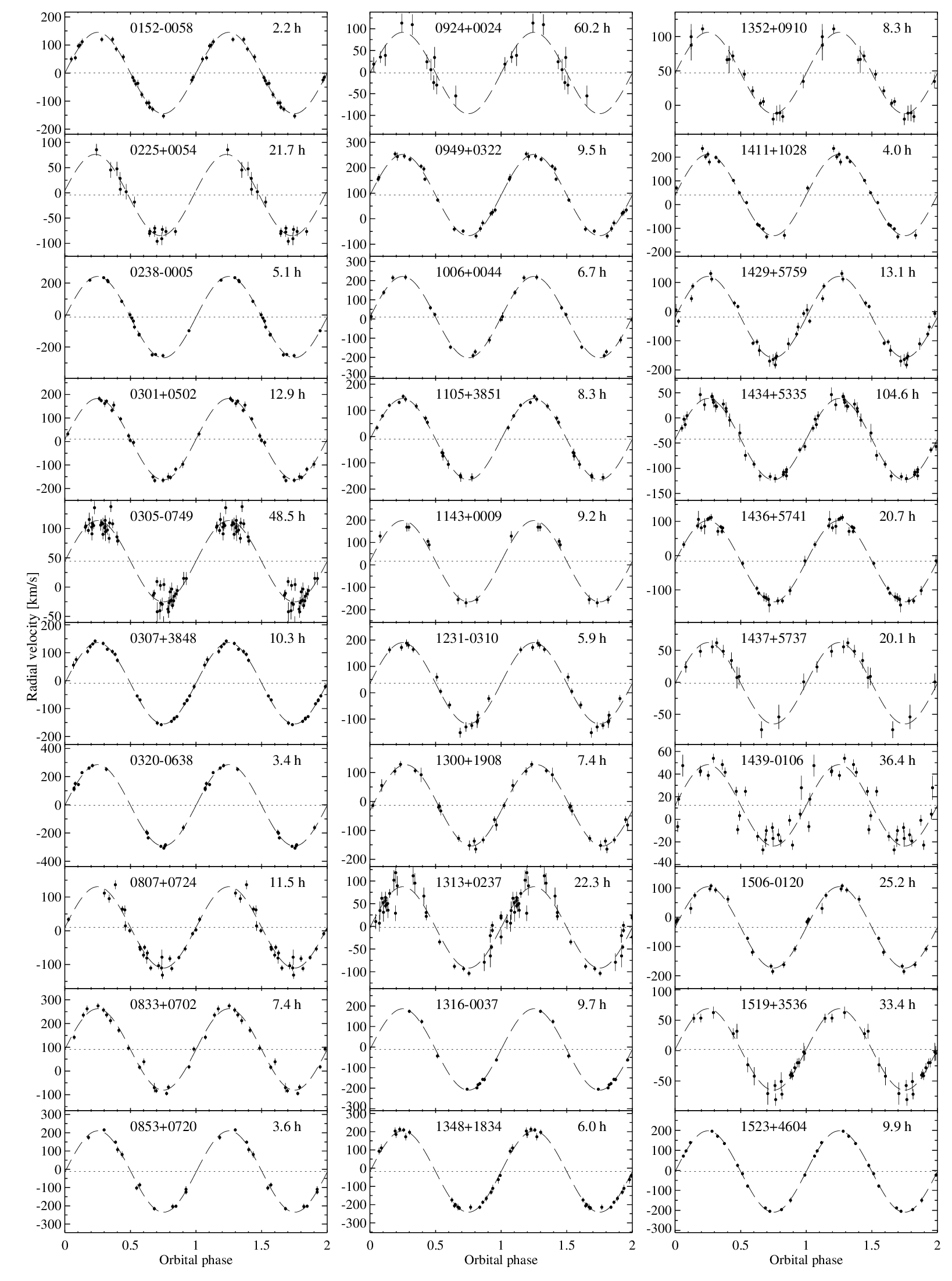}
\caption [Radial velocities]{Phase folded radial velocities curves and
sine  fits to the  data (dashed lines). As for Fig.~\ref{g:spectra1}
systems are sorted in right ascension  from left to right and from top
to bottom.}
\label{g:rv1}
\end{center}
\end{figure*}
}%
\onlfig{5}{
\begin{figure*}
\begin{center}
\includegraphics[angle=0,clip=,height=\textheight]{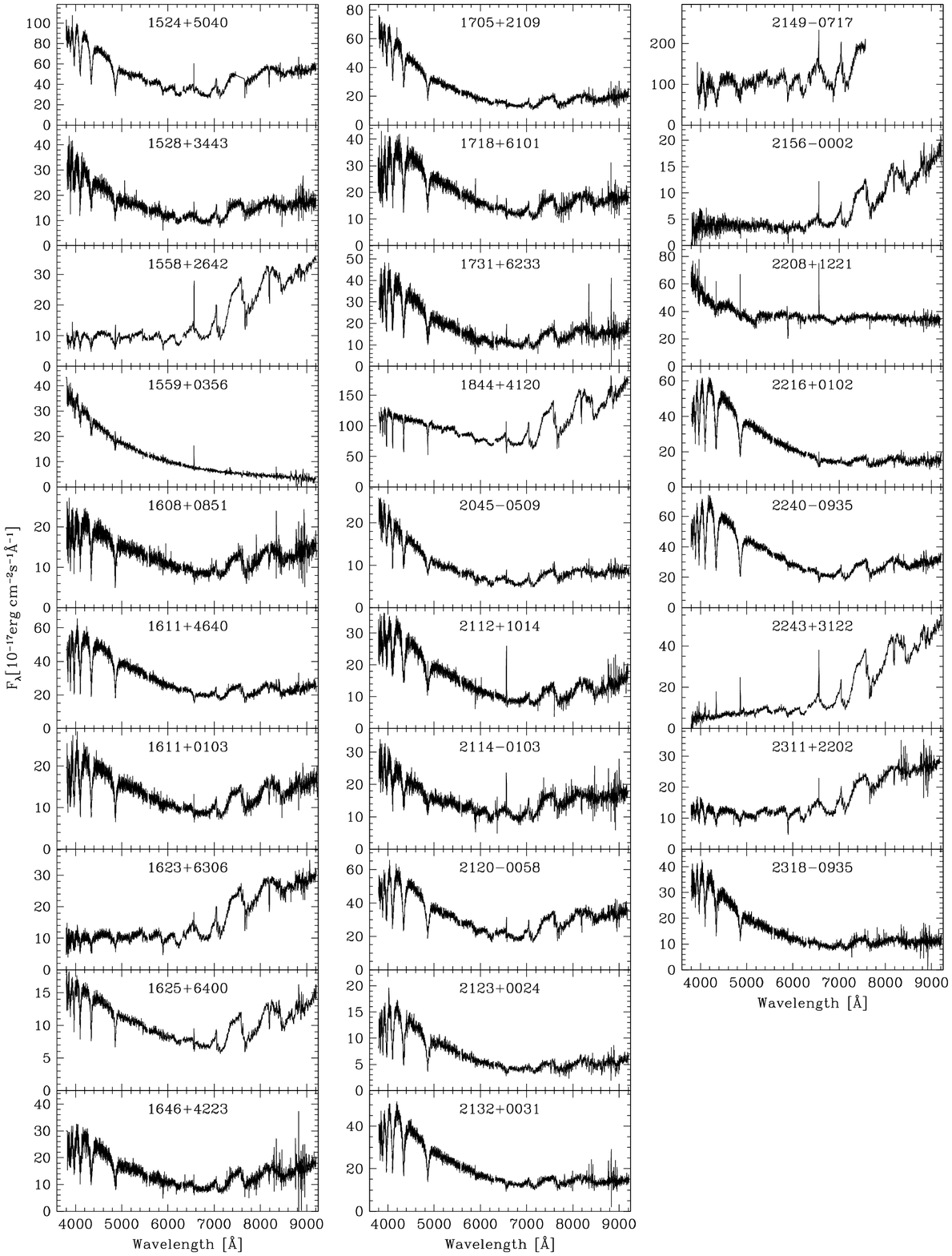}
\caption [SDSS  spectra]{SDSS spectra of  the 58 systems  with orbital
period measurement. Systems are sorted in right ascension from left to
right and from top to bottom.}\label{g:spectra2}
\end{center}
\end{figure*}
}%
\onlfig{6}{
\begin{figure*}
\begin{center}
\includegraphics[angle=0,clip=,height=\textheight]{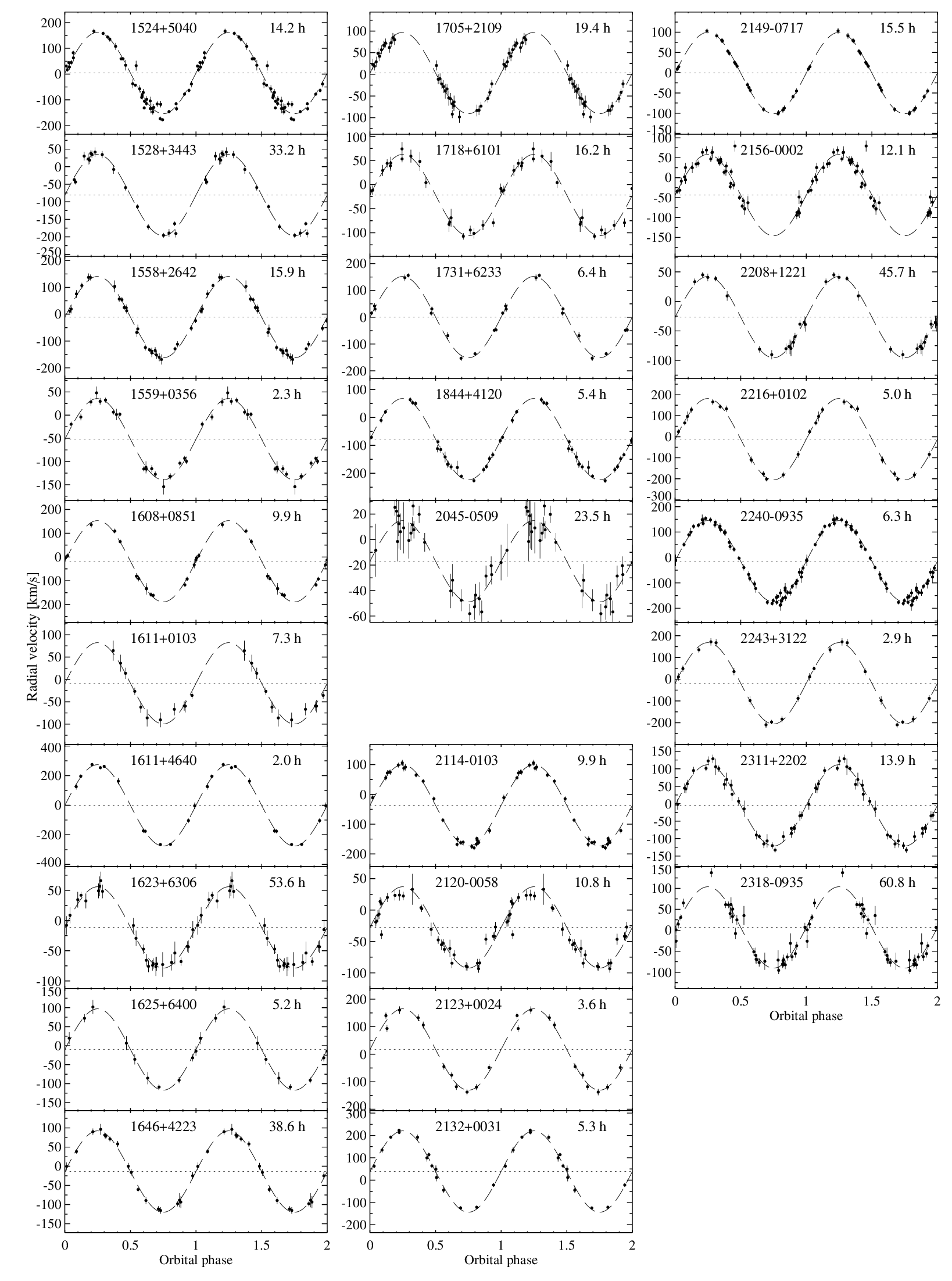}
\caption [Radial velocities]{Phase folded radial velocities curves and
sine  fits to the  data (dashed lines). As for Figure~\ref{g:spectra1}
systems are sorted in right ascension  from left to right and from top
to bottom.}\label{g:rv2}
\end{center}
\end{figure*}
}%
\subsubsection{Orbital inclinations and candidate eclipsing systems}
\label{sec:incl}

We estimated the orbital inclinations and binary separations of those
PCEBs for which masses of the white dwarf and the companion star (Sect.~\ref{sec:sample}),
$K_2$ and $\Porb$ were available using Kepler's third law, and
propagated the errors in all quantities accordingly. We further
calculated $K_\mathrm{WD}$, as well as the Roche-lobe size of the
companion star, $R_\mathrm{lob}$ using \citeauthor{eggleton83-1}'s
(1983) approximation. Finally, using the empirical radius-spectral
type relation of \cite{rebassa-mansergasetal07-1} we estimate the
Roche-lobe filling factors of the secondary stars. All this ancillary
information is given in Table~\ref{t:allparam}. In case the effective
temperature of the primary is smaller than $12~000$~K, the derived
spectroscopic mass is inherently uncertain \citep{koesteretal09-1,
  tremblayetal10-1} and the derived parameters should be used with
caution.

For six systems, the estimated orbital in\-cli\-na\-tions are clo\-se to 90 
degrees, making them good candidates for being eclipsing binaries:
SDSS\,0238--0005, SDSS\,1348+1834, SDSS\,1434+5335, SDSS\,1506--0120,
SDSS\,2132+0031 and SDSS\,2318--0935. Eclipsing systems offer the best
opportunity to derive accurate fundamental stellar parameters,
i.e. masses and radii, and we have begun to exploit the potential
offered by the eclipsing PCEBs \citep{steinfadtetal08-1, nebotgomez-moranetal09-1,
  pyrzasetal09-1}. SDSS\,1348+1834, aka CSS\,21357, was recently
identified in the Catalina Real Time Transient Survey
\citep{drakeetal09-1} as an eclipsing PCEB \citep{drakeetal10-1},
confirming the predictive power of our estimates.

\begin{figure}[!t] 
\begin{center}
\includegraphics[angle=90,clip=,width=0.5\textwidth]{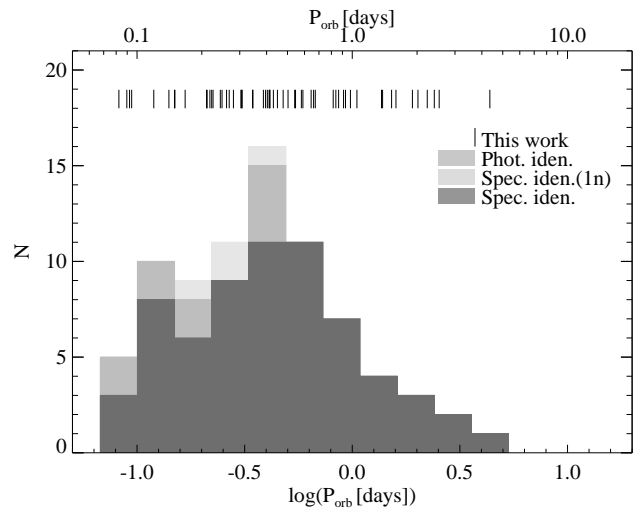}
\caption{Orbital period distribution of the 79 SDSS PCEBs characterised to date.
  Systems identified as PCEBs based on spectroscopic
  observations are shown in light gray if the spectra were taken during the 
  same night and dark gray otherwise. In medium gray we show those systems
  that were identified as PCEBs based on photometry. The 58 systems
  added by this paper to the orbital period distribution are indicated
  with vertical lines. \label{g:porbs58}}
\end{center}
\end{figure}

\subsubsection{Observed orbital period distribution}
\label{sec:obspdist}

We have measured here the orbital periods of 58 PCEBs.  Together with
the 15 periods already published by our group
\citep{pyrzasetal11-1,rebassa-mansergasetal08-1, schreiberetal08-1, pyrzasetal09-1,
  nebotgomez-moranetal09-1}, two more by \cite{drakeetal10-1}, and
four additional PCEBs that are subject to forthcoming individual
studies (Pyrzas et al in prep.; Parsons et al in prep.), the total
number of PCEBs from SDSS with orbital period measurements stands at
79, which is more than double the corresponding number of pre-SDSS
PCEBs.  The orbital period distribution of the SDSS PCEBs is shown in
Fig.~\ref{g:porbs58} on a logarithmic scale, in light and dark gray
those systems that were identified as PCEBs based on spectroscopy
(separating systems with PCEB identification spectra taken in the same
night, light gray) and photometrically identified PCEBs are shown in 
medium gray. 
The 58 periods determined in this paper are indicated by
vertical lines. Taking this observed period distribution at face
value, PCEBs are found to have orbital periods ranging from 
1.97\,h to 4.35\,d, they follow approximately a normal distribution 
in $\log(\Porb)$ with a peak centred on $\sim8.1$\,h and a width of 
$0.41$ in $\log(\Porb[d])$ as determined from a Gaussian fit.

\section{Analysis} 
\label{sec:porb_dis}

We have performed a radial velocity survey of 1246 WDMS binaries. The
majority of the systems have three radial velo\-ci\-ty measurements,
spread over hours to hundreds of days (see Fig.\,\ref{g:num_spec}). We
have identified 191 of these systems as PCEBs based on the detection
of significant radial velocity variations. While we consider the
remaining 1055 systems to be candidate wide binaries, it is clear that
there will still be a number of PCEBs among them which our radial
velocity survey failed to identify because of observational selection
effects.

Our ultimate aim is to establish the intrinsic period distribution of
the SDSS PCEBs. We will first investigate in the following sections
what information might be gleaned from the distribution of probabilities 
of our initial radial velocity survey
(Fig.\,\ref{g:probab}). Subsequently, we will analyse the
observational biases within the orbital period distribution of the
SDSS PCEBs (Fig.\,\ref{g:porbs58}), and derive a bias-corrected
period distribution, that will then be discussed in
Sect.\,\ref{sec:discussion} in the context of compact binary
evolution.

\subsection{Does our radial velocity survey constrain the 
orbital period distribution of WDMS binaries?}

\begin{figure}[!t] 
\begin{center}
\includegraphics[angle=-90,width=\columnwidth]{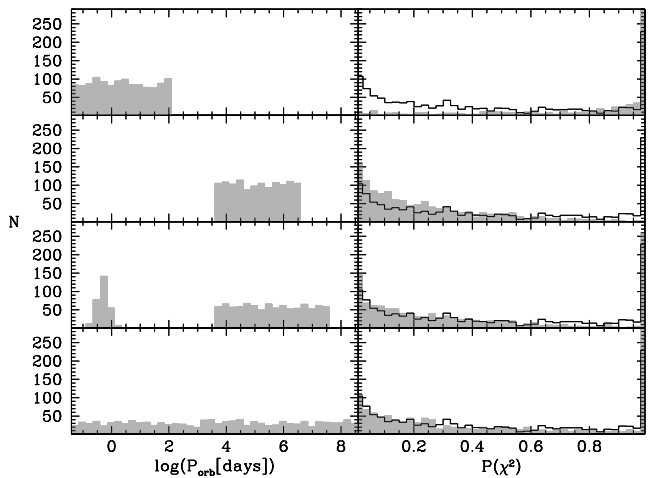}
\caption   [Orbital  period   distribution]{Assumed   orbital  period
distributions (left  panels) and associated  simulated distribution of
the probability  of measuring large radial  velo\-ci\-ty variations (gray)
with   measured   distribution   in   black  for   comparison   (right
panels). }\label{g:SDSS_Porbs}
\end{center}
\end{figure}

The detailed shape of the probability distribution used to identify
PCEBs among SDSS WDMS binaries will depend on the technical
details of our radial velocity survey (spectral resolution, temporal
sampling, number of spectra) and the intrinsic period distribution of
the WDMS binaries. To test how sensitive the observed probability
distribution is to differences in the underlying orbital period
distribution, we performed extensive Monte Carlo simulations. As input,
we use the measured stellar masses (Table\,\ref{t:allparam}), the
times of observation, and the radial velocities and associated errors
(Table\,\ref{t:rvs_iden}) for all 1246 WDMS
binaries within our radial velocity survey. We then adopted four
diffe\-rent orbital period distributions: (1) short-period binaries
only, with $2\,\mathrm{h}<\Porb<100$\,d, and uniformly distributed
in $\log(\Porb)$; (2) wide binaries only, with
$3.5<\log(\Porb[\mathrm{d}])<6.5$, uniformly distributed in
$\log(\Porb)$; (3) a bi-modal distribution with a short-period binary
population peaking at 7.5\,h, normally distributed in $\log(\Porb)$
and with a minimum period of $\sim 1.5$\,h, plus a second population
of wide binaries with $3.5<\log(\Porb[\mathrm{d}])<7.5$, uniformly
distributed in $\log(\Porb)$, and finally (4) a uniform period
distribution in $\log(\Porb)$ with
$\log(\Porb[\mathrm{d}])<8.5$. Obviously (1), (2) and (4) are very
unlikely to be physically relevant, but were adopted to run the test
with several radically different period distributions. We then used $10^4$
random inclinations per object, uniformly distributed in $\cos i$, 
and calculated the probability distributions of the two adopted period
distributions. We assumed circular orbits for simplicity (and note 
that the orbits of PCEBs will be, in most cases, circularised).

The results of this simulation are illustrated in
Fig.\,\ref{g:SDSS_Porbs}, showing the adopted period distributions on
the left hand side, and the resulting probability distributions on the
right hand side, with the observed probability distribution shown for
comparison. The top two panels show that a peak at $P(\chi^2)=1$
requires close binaries, whereas a broad accumulation towards
$P(\chi^2)=0$ is domi\-na\-ted by wide binaries. In other words, the
observed pro\-ba\-bi\-li\-ty distribution (Fig.\,\ref{g:probab}) can only be
explained if the underlying orbital period distribution of the
observed WDMS binaries contains both wide binaries
($\Porb\gappr10^3$\,d) and PCEBs with shorter periods
($\Porb\lappr\,100$\,d). 

However, the bottom two panels show that the 
two period distributions which contain both short-period PCEBs and
wide binaries, but radically differ in shape, both lead to probabili\-ty
distributions that are very similar to the observed one. We conclude
that our radial velocity survey carried out to identify PCEBs alone
cannot constrain the detailed shape of the underlying WDMS binary orbital 
period distribution, and that it is hence necessary to accurately
measure the orbital periods for a large and homogeneous sample of
PCEBs.

\subsection{Observational biases and selection effects}
\label{sec:biases}

We have presented in Sect.\,\ref{sec:obspdist} the orbital period
distribution of 79 PCEBs that were spectroscopically identified by
SDSS (Fig.\,\ref{g:porbs58}). The resulting orbital period
distribution peaks at $\sim8$\,hrs and very few systems with orbital
periods longer than a day have been found.

This is not only the largest but also the most homogeneous sample of
PCEBs available so far, which has a large potential to provide crucial
constraints on the theories of close compact binary formation and
evolution \citep[e.g.][]{schreiberetal10-1, zorotovicetal10-1,
  rebassa-mansergasetal11-1}. It is, however, of fundamental
importance to understand any observational selection effects that may
affect the properties of the SDSS PCEB sample. In the context of the
orbital period distribution, there are two particular points that need
to be investigated. Firstly, our radial velocity PCEB identification
method is obvious\-ly more sensitive to short orbital period and high
inclination systems and secondly, we only measured the orbital periods
of a sub-sample of the identified PCEBs which is not necessarily
representative for the entire sample.

\begin{figure}
\begin{center}
\includegraphics[angle=-90,clip=,width=0.5\textwidth]{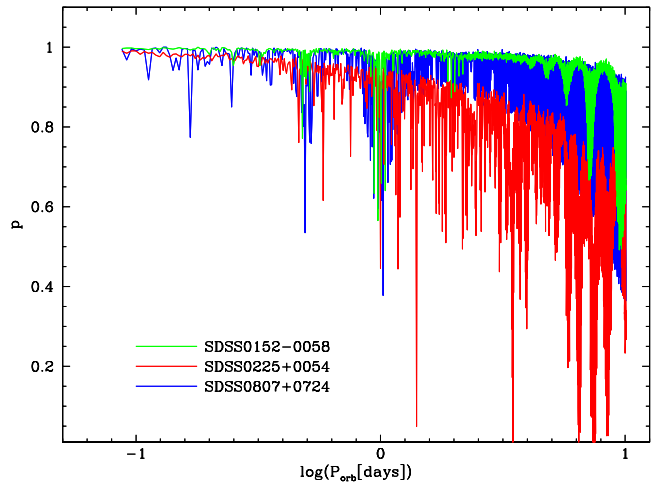}
\caption []{Detection  probability for three cases  corresponding to 2
spectra taken at the VLT (green), 6 spectra taken at the WHT, Magellan
and SDSS (blue) and 16 spectra from the SDSS (red). \label{g:biases2}}
\end{center}
\end{figure}

\subsubsection{PCEB identification}
\label{sec:pceb_iden}
While we have compiled the largest and most homogeneous
sample of PCEBs to date, not all the systems in this sample have been
identified in the same way.  

Ten systems were initially classified as PCEBs because of photome\-tric
variability \footnote{SDSS\,0106--0014, 0110+1326, 0303+0054,
  0908+0604, 1212--0123, 1423+2409, 1548+4057, 1648+2811, 1724+5620 and
  2112+1014} and four because of radial velocity variations seen over
the course of a single night\footnote{SDSS\,0307+3848, 1300+1908,
  1348+1834 and 1625+6400}. Both methods favour the
identification of short orbital period PCEBs. This is illustrated in
Fig.~\ref{g:porbs58}, where all those 14 systems have periods
$\la9$\,h (0.4\,d), and we discard them in the following analysis.

The remaining 65 PCEBs were identified by radial veloci\-ty variations
from spectra that have been obtained on different nights, and we will
concentrate our analysis of the observational biases on this ``core''
sample. 
The strength of the observational bias introduced in a radial
velocity search for binaries depends on the intrinsic orbital period
and the sampling frequency.

We have performed Monte-Carlo simulations similar to those presented
in \citet{schreiberetal08-1, rebassa-mansergasetal08-1,
  schreiberetal10-1} to evaluate if the observed orbital period
distribution (Sect.~\ref{sec:Porb}) is significantly biased by the
design of our radial velocity survey, and we calculated the detection
probability as a function of orbital period for all 65 PCEBs.  Given
the times of observation and the error of the radial velocity
measurements, we randomly select $10^4$ phases and inclinations for a
given orbital period, and determine the co\-rres\-pon\-ding $\chi^2$
probability. The PCEB detection probability for each system as a
function of the orbital period is then simply given by the number of
cases with $P(\chi^2)>0.9973$ ($>3\sigma$) divided by $10^4$.  The resulting
detection probabilities depend on the number of spectra taken, the
temporal sampling of the measurements, and the accuracy of the radial
velocity measurement (i.e. basically the spectral resolution of the
telescope/instrument combination that has been used).

\begin{figure}[!t]
\begin{center}
\includegraphics[angle=90,clip=,width=0.5\textwidth]{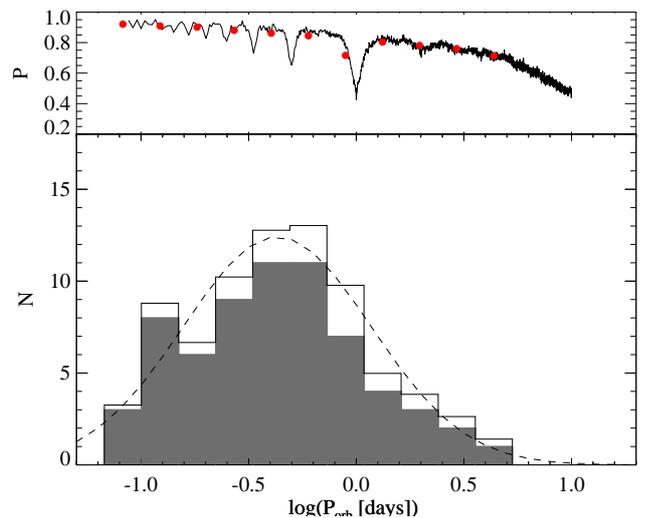}
\caption []{Orbital period distribution of the 65 SDSS PCEBs
  discovered in an homogeneous way through radial velocity variations
  measured in spectra spread over different nights (gray) and the 
  bias-corrected 
  distribution (white). The Gaussian fit to the bias-corrected distribution
  is plotted with a dashed line. In the upper panel we show the mean
  probability of detection as a function of the orbital period (see
  text) and the corresponding mean value at each bin centre (red
  dots).\label{g:biases}}
\end{center}
\end{figure}

Figure~\ref{g:biases2} shows the obtained detection probabilities for
three rather extreme examples: SDSS\,0807+0724 (green,
$\Porb=11.45$\,h, identified as a PCEB from two VLT/FORS spectra),
SDSS\,0152--0058 (blue, $\Porb=2.15$\,h, $3\times$WHT,
2$\times$Magellan, 1$\times$SDSS), and SDSS\,0225+0055 (red,
$\Porb=21.86$\,h, 16 SDSS subspectra). We have chosen these three
cases to illustrate that our PCEB identification method was sensitive
to PCEBs with orbital periods in the range of $1-10$\,days (detection
probabilities $P(\chi^2)\gappr0.35$) as long as the time baseline exceeded
one night.  In addition, Fig.~\ref{g:biases2} illustrates the
following: (1) the lower the re\-so\-lu\-tion, the lower is the detection
probability, even in the most favourable case, i.e. a large number of
spectra.  (2) Variations in the detection probability with orbital
period are dominated by phase coverage and are therefore strongest
when only few spectra are available. Orbital periods near one day 
and their first subharmonics (12\,h, 8\,h) have substantially depressed
detection probabilities, as this increases the chances to sample the
same orbital phase twice on two subsequent nights. (3) Superimposed on
the fine structure related to the details of the sampling is a
decrea\-sing envelope, which is caused by a larger range of (low)
inclinations leading to radial velocity variations that are too low
to be resolved. The mean detection probability for the whole sample,
calculated from the 65 individual detection probabilities, is
higher than $35\%$ up to 10 days orbital periods
(Fig.\,\ref{g:biases}, upper panel), demonstrating that our radial
velocity survey is well capable of identifying PCEBs with long orbital
periods. We will use this mean detection probability to bias-correct
the observed orbital period distribution in Sect.~\ref{sec:porbbiascor}.

\subsubsection{PCEBs selected for orbital period measurements}
\label{sec:biasporb}

\begin{figure}[!t]
\begin{center}
\includegraphics[angle=90,clip=,width=0.5\textwidth]{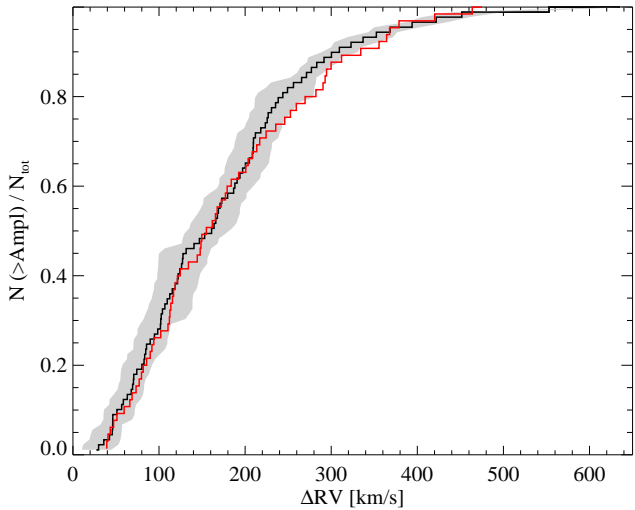}
\caption []{Cumulative distribution function (CDF) of the radial velocity
  variations used for the identification of PCEBs, systems with
  measured orbital period are shown in red and systems that were not
  further followed-up are shown in black. The gray area is the 
  corresponding CDF taking the $1\sigma$ radial velocity errors into account).  
  \label{g:cdfs_rv}}
\end{center}
\end{figure}

Given that we have measured orbital periods only for a fraction of the
PCEBs that we identified, the next question to address is whether the
subset of systems is representative of the whole sample. For instance,
our follow-up strategy might have been biased towards the ``easier''
targets, i.e. those which exhibited large variations in our radial
velocity survey. For this purpose, we compared the cumulative radial
velocity variations of the 65 PCEBs identified in our radial velocity
survey (see Sect~\ref{sec:subspectra_ana}), and of those PCEBs 
for which we obtained no additional follow-up
observations (Fig.\,\ref{g:cdfs_rv}). A Kolmogorov-Smirnov
test between both sets gives a probability of 0.97 of both samples
being drawn from the same parent population.  
We can conclude that both distributions are very similar, and
that we have measured the orbital period of a representative
sub-sample of the SDSS PCEB population.

\subsubsection{Bias corrected orbital period distribution}
\label{sec:porbbiascor}

We have demonstrated in Sect.\,\ref{sec:biasporb} that the 65 PCEBs
that were identified from multiple-night radial velocity snapshots and
have measured orbital periods are representative of the entire SDSS
PCEB sample. Here, we apply the average detection probability
(Sect.\,\ref{sec:pceb_iden}) to correct their observed period
distribution (dark gray histogram in Fig.~\ref{g:biases}). 
The upper panel of Fig.\,\ref{g:biases} shows the detection probability
averaged across all 65 systems, and its binned values as red dots.
The bias-corrected period distribution is shown in
Fig.\,\ref{g:biases} (white histogram). It is evident that the
observational bias of our radial velocity survey does not dramatically affect 
the conclusions drawn in Sect.\,\ref{sec:obspdist}, i.e. the period
distribution of the SDSS PCEB sample follows broadly a normal
distribution centred at $\sim10.3$\,h (0.43\,d), a width of $0.44$ 
in $\log(\Porb[d])$, a short-period cut-off at $\sim2$\,h and a long-period cut-off at
$\sim5$\,d, where the position of the peak and the width have been 
determined from a Gaussian fit (dashed line).
The bias-corrected period distribution is
discussed in the context of published binary population models in
Sect.\,\ref{sec:comp_bps}.

\section{Discussion}
\label{sec:discussion}

\subsection{The PCEB fraction among WDMS binaries}
\label{sec:pcebfraction}
Analysing the radial velocity variations of 1246 SDSS WDMS binaries we
identified 191 PCEBs (Sect.~\ref{sec:subspectra_ana}), which provides a lower
limit of $\sim15\pm1$\% to the fraction of PCEBs among the SDSS WDMS
binaries. However, for a substantial number of these 1246 systems the
only available source of radial velocities are the SDSS subspectra,
which, as they are usually taken during a single night (see
Fig.\ref{g:num_spec}), are sensitive only to very short orbital
periods. 

If we only consider the sub-sample of 718 WDMS binaries that were
observed on different nights, we find 164 PCEBs, i.e. a PCEB fraction
of $\sim 23\pm2$\%. This number is similar to that found by
\cite{schreiberetal08-1}, based on our VLT/FORS radial velocity survey
of 26 SDSS WDMS binaries. This empirical determination of the PCEB
fraction is broadly consistent with the results of the binary
population synthesis carried out by \cite{willems+kolb04-1}. 

Two observational selection effects will affect this empirical
estimate of the PCEB fraction. On the one hand, our analysis in
Sect.\,\ref{sec:pceb_iden} showed that a number of PCEBs will escape
identification because of their low orbital inclination. 
From the mean detection probability and the bias corrected orbital period 
distribution 
we learn that $\sim16\%$ of the PCEBs are not detected due 
to a combination of inclination and phase coverage (see Fig.~\ref{g:biases}).
 Assuming that the 164 detected PCEBs 
(observed in diffe\-rent nights) follow the same orbital period distribution 
as the observed one (drawn from 65 of them) the number of expected PCEBs is 
195, implying
a fraction of PCEB to WDMS binaries of $27\pm2\%$ ($195/718$).
On the other hand, WDMS binaries that are sufficiently wide will appear as 
blended objects, or even fully resolved objects, and not have been followed up
spectroscopically, as they have unreliable colours or colours of
single white dwarfs and M-dwarfs, respectively. 
Hence, our estimate of $27\pm2\%$ PCEBs among the spatially 
unresolved SDSS WDMS binaries can be taken as an
upper limit on the intrinsic fraction 
of PCEBs among all WDMS. 
\citet{schreiberetal10-1} estimated the fraction of SDSS resolved WDMS to be 
12-23\%. Taking this into account we finally estimate the intrinsic fraction of
PCEBs to be $\sim21-24$\%.

\citet{farihietal06-1, farihietal10-1} pursued a complementary
approach to investigate the make-up of the known WDMS binary
population, obtaining high spatial resolution imaging with
\textit{HST}/ACS. They observed a total of 72 WDMS binaries unresolved
from the ground, and found a bimodal distribution in projected binary
separations: all spatially resolved systems had binary separations
$\ga10$\,AU, clearly above the detection threshold of
$1-2$\,AU. Accounting for the frequency of WDMS binaries that can be
resolved from the ground \citet{farihietal05-1,farihietal10-1} 
conclude that the fraction of short-period
binaries among all WDMS binaries is $\simeq25$\%, which is in
excellent agreement with our estimate presented here. 19 of the WDMS
binaries imaged with HST were also spectroscopically re-discovered by
SDSS, and are hence in our WDMS binary sample (Table\,\ref{t:farihi}).

\citet{farihietal10-1} predict that all the WDMS binaries unresolved
by \textit{HST} (Table\,5 in their paper) should be short-period
PCEBs, which is confirmed for all five systems for which we have
obtained follow-up observation, spanning more than one night. For an
additional two of Farihi's unresolved systems (SDSS1341+6026, 
1500+1659) we have only the radial velocity information of
the SDSS subspectra spanning less than one hour, and we classified both
systems as wide binary candidates. This underlines what we said
above, i.e. that the single-night SDSS subspectra allow us to identify
only the shortest-period WDMS binaries (which we fully take into
account by our Monte Carlo analysis of the selection biases). For
completeness, we note that one additional unresolved binary from
\citet{farihietal10-1} was recently found to be an eclipsing WDMS
binary with an orbital period of 2.93\,h \citet{drakeetal10-1}.

Vice versa, based on our radial velocity information for nine of the
WDMS binaries spatially resolved by \textit{HST} (Table\,4 of
\citealt{farihietal10-1}), we classified all of them as wide binary
candidates (labeled as ``wide'' in the bottom portion of 
Table~\ref{t:farihi}).

\input{table6.tex}

\subsection{PCEBs and detached CVs in the orbital period gap}
The observed orbital period distribution of CVs presents a sharp
decrease of systems between 2--3\,h \citep{warner76-1,whyte+eggleton80-1,
  knigge06-1}, referred to as the orbital period gap of CVs, which is
interpreted as a key to understanding their long-term evolution.  The
mass ratio of CVs implies that they are undergoing stable mass
transfer, and it is angular momentum loss that drives them towards
shorter orbital periods, keeping the companion stars in Roche-lobe
contact. Above the period gap, the secondary stars have a radiative
core, and are subject to magnetic wind braking that acts as a
strong agent of angular momentum loss. The corresponding mass loss
rates are sufficient to drive the companion stars of these long-period
CVs out of thermal equilibrium. At $\sim3$\,h, the companion stars
become fully convective, and magnetic wind braking is thought to
stop. From there on, the evolution towards shorter periods is driven
by gravitational wave radiation, resulting in much lower mass loss
rates. Consequently, the secondary stars have time to relax to their
thermal equilibrium radii, shrink within their Roche lobes, and turn
into detached WDMS binaries with no mass transfer. Gravitational wave
radiation keeps driving these detached systems towards shorter
periods. At $\sim2$\,h the secondary stars re-establish Roche-lobe
contact, mass transfer stars again, and the systems re-appear as
CVs. This ``disrupted magnetic braking'' scenario is currently the
standard paradigm of CV evolution \citep{rappaportetal83-1}.

Observationally, there is very little difference between ge\-nuine
pre-CVs that will start mass transfer within, or shortly below, the CV
period gap, and those detached CVs (dCVs) that are crossing the
gap. The common hallmark of both populations is that they are WDMS
binaries with secondary spectral types $\sim$M6 to $\sim$M4. A very
simple, but so far untested, prediction of the disrupted magnetic
braking scenario is that there should be an \textit{excess} of systems
with periods of 2--3\,h in the period distribution of WDMS binaries, 
due to dCVs that are crossing the orbital period gap.
\cite{davisetal08-1} carried out a number of binary population studies
to estimate the strength of this effect, and predict that within the
CV period gap the number of dCVs should exceed that of genuine pre-CVs
with secondary stars in the mass range $0.17-0.36$\Msun\, by a factor
$\sim4-13$, which implies an increase of systems (PCEBs plus dCVs) 
of 1.5--2 in the gap in comparison with PCEBs at longer periods 
(see their Figs.\,4 \&10).

\begin{figure}[t!]
\begin{center}
\includegraphics[angle=90,width=0.5\textwidth]{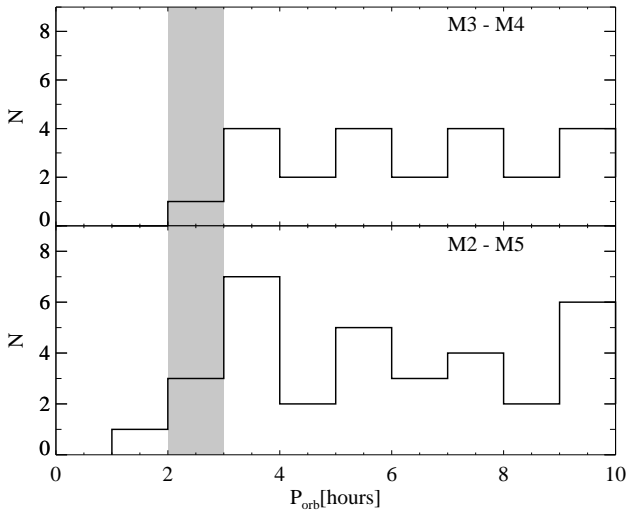}
\caption[]{Orbital period distribution of PCEBs with secondary
  spectral types in the range M3--M4 (upper panel) and M2--M5 (bottom
  panel).  Magnetic braking disruption predicts a peak at around 2--3
  hours corresponding to an increase in systems due to detached CVs,
  we highlighted the region in gray.}\label{g:porb_gap}
\end{center}
\end{figure}

The large sample of SDSS PCEBs provides a first opportunity to test
this prediction. Observationally, the secondary stars in the SDSS WDMS
binaries are characterised in terms of their spectral type. CVs just
above and below the period gap have $\sim$M3--4, and $\sim$M5
secondary stars, respectively. We show in Fig.\,\ref{g:porb_gap} the
period distributions of all SDSS WDMS binaries with M3--4 companions
(top panel), and a somewhat broader range of M2--5 (bottom panel)
which accounts, somewhat conservatively, for the uncertainty in the
spectral types of the companion stars.  
The expected increase towards shorter periods by a factor of 1.5--2
is not present in our data. In contrast the number of dCVs plus PCEBs is 
slightly decreasing (M3--4) or constant (M2--5), although the number of systems 
in this sub-sample is still rather small. Comparing the number of 
dCVs plus PCEBs in the gap with the number of PCEBs between 3--4 hours, 
we obtain a fraction of dCVs plus PCEBs (in the gap) to PCEBs (above the gap) 
of $0.25\pm0.30$ (M3--4) or
$0.43\pm0.30$ (M2--5), which excludes the predicted increase of 1.5--2 on a 
$3\sigma$ (M3--4) or a $2\sigma$ (M2--5) level.

Three SDSS PCEBs that match the expected characte\-ris\-tics of dCVs are
SDSS\,0052--0053, SDSS\,1611+4640 and, SDSS\,2243+3122, with orbital
periods 2.74, 1.98 and 2.87\,h res\-pec\-tively.  The Roche-lobe filling
factors of SDSS\,0052--0053 and SDSS\,1611+4640 are $1.0 \pm 0.6$ and
$1.52 \pm 0.50$ respectively. For SDSS\,2243+3122 we have no estimate of 
the (DC) white dwarf mass, and hence can not determine the filling factor. 

\subsection{Comparison with non SDSS PCEBs}
\begin{figure}[!t] 
\begin{center}
\includegraphics[angle=90,clip=,width=0.5\textwidth]{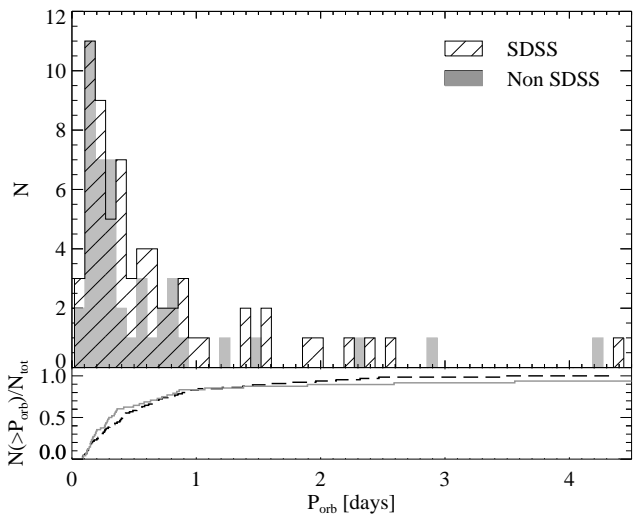}
\caption [ Orbital period distribution]{Orbital period distribution of
the 48 non-SDSS PCEBs that contain  a white dwarf plus a main sequence
star (gray histogram) and the 65 systems SDSS PCEBs discovered through
radial  velocity  variations  on  spectra taken  on  different  nights
(dashed  histogram)  in days.  Bottom  panel: cumulative  distribution
function of  non SDSS PCEBs (gray  line) and SDSS PCEBs  (black dashed
line). }\label{g:Porbs}
\end{center}
\end{figure}

In this section we compare the orbital period distribution of PCEBs
discovered within the SDSS, with that of PCEBs that were discovered by
other surveys. We compiled a list of 48 non-SDSS PCEBs containing a
white dwarf plus a main sequence star from
\citeauthor{ritter+kolb03-1} (\citeyear{ritter+kolb03-1}, version
v7.14) and \cite{brownetal11-1}.  The orbital period distribution of
non-SDSS PCEBs (Figure~\ref{g:Porbs}, gray histogram) is overall
similar to that of the SDSS PCEBs (white hatched histogram), presenting
both a peak at around $\sim5$\,h and decreasing number of systems towards
long orbital periods. A Kolmogorov-Smirnov test between the cumulative
period distributions (bottom panel in Figure~\ref{g:Porbs}) gives a
probability of 35\%, i.e. consistent with the two samples being drawn
from the same parent population.

We note that this result is in agreement with the conclusions of
\cite{miszalskietal09-1}, who first noticed the resemblance 
between the orbital period distribution of binary central stars in 
planetary nebulae (PNe) and that one of PCEBs as presented by 
\citet{rebassa-mansergasetal08-1}. This was later confirmed by 
\cite{hillwig11-1}, 
who selected 27 WDMS binaries that are the central
stars of planetary nebulae (PNe) and the 35 SDSS WDMS binaries (as
published by \citet{zorotovicetal10-1}). 
Surveys dedicated to measure the orbital period of the binary central 
stars of PNe have been mostly photometric 
\citep[see][and references therein]{bond00-1}, implying a bias towards 
short orbital periods. \cite{demarcoetal08-1}
studied the possible biases introduced in the photometric sample, and
concluded that the decrease of the number of  PNe towards long periods 
is unlikely to be due to observational biases. 

\subsection{Comparison with predictions from BPS models}
\label{sec:comp_bps}

In this section we discuss the SDSS WDMS binary sample in the context
of published binary population synthesis models. However, it is 
necesary to keep in mind that the SDSS PCEB sample is subject to selection 
effects, i.e. systems with com\-pa\-nion spectral types earlier than 
$\sim$M0 are severely underrepresented, and a comparison of the predicted 
orbital period distribution of the different BPS studies for the 
corresponding mass domain is in most cases not possible.

Common to all BPS
models is the prediction that the population of WDMS binaries should
be made up of wide binaries, with orbital periods longer than one year, 
in which the components never interacted, and close binaries which formed from
common envelope evolution. The two most important parameters
determining the outcome of the CE phase are the initial mass ratio
distribution of the progenitor binaries, and the efficiency of the
ejection of the common envelope, $\alpha_\mathrm{CE}$, which will
affect the position of the peak in the orbital period distribution of
the PCEBs, and the slope towards longer orbital periods.

One of the first BPS for WDMS binaries was carried out by
\cite{dekool92-1} and \cite{dekool+ritter93-1}, predicting that most
PCEBs should have periods shorter than $\sim$~one day.  An extended
tail towards longer orbital periods up to 1000 days is expected for
$\alpha_\mathrm{CE}=1$, while no tail is observed for
$\alpha_\mathrm{CE}=0.3$. The authors also simulated the expected
properties of a volume limited blue sample of present-day WDMS
binaries, however, comparison with the very small number of PCEBs at
that time did not lead to any significant observational
constraints. Our orbital period distribution (Fig.\,~\ref{g:biases})
is very similar to their predicted one for $\alpha_\mathrm{CE}=0.3$
and the mass of the secondary star picked randomly from the IMF, 
already pointed out by \cite{miszalskietal09-1} in the context of PN, 
or with $dN\propto dq$ (see their Fig.\,12b). However, this comparison
should be regarded with some caution, as \cite{dekool+ritter93-1}
simulated their ``observational'' sample for the selection criteria of
the Palomar-Green Survey \citep{greenetal86-1} ($U<16$ and
$U-B<-0.5$), whereas the SDSS WDMS sample has a much larger limiting
magnitude $i\simeq19.1$ and is drawn from a much larger $ugriz$ colour
volume.

\cite{willems+kolb04-1} carried out an extensive BPS study of WDMS
binaries, using $\alpha_\mathrm{CE}=1$, and 
for different initial mass ratio distributions. 
Their BPS predicts
that the orbital period distribution of PCEBs peaks around 1~day, with
a substantial tail towards longer orbital periods of up to 100 days,
which is in contrast to the period distribution of the SDSS PCEBs
(Fig.\,~\ref{g:biases}). However, \cite{willems+kolb04-1} did not
convolve their model WDMS binary population with detailed
observational selection effects, but just discussed their results in
terms of luminosity ratio distributions,
$L_\mathrm{wd}/L_\mathrm{MS}$, which prevents any more quantitative
comparison with our SDSS WDMS binary sample.

\cite{politano+weiler07-1} investigated the effects on the po\-pu\-la\-tion
of PCEBs for $\alpha_\mathrm{CE}$ depending on the mass of the
se\-con\-da\-ry star. They predict an orbital period distribution peaking at
around 3 days, 
which is incompatible with the observed period distribution of the SDSS WDMS
binaries. However, also \citet{politano+weiler07-1} did not simulate
a realistic ``observational'' WDMS population, again preventing a
quantitative comparison with our sample. 

The most recent BPS of WDMS binaries was performed by
\cite{davisetal10-1}, who, in contrast with previous works, 
calculate the binding energy parameter $\lambda$ from stellar
evolution models incorporating the internal energy of the
envelope. While this is a significant step forward compared to
previous PBS that worked with a constant value of $\lambda$,
\cite{davisetal10-1} obtained the values of
$\lambda$ from the tables of \cite{dewi+tauris00-1}, 
which cover only a very small part of the relevant parameter
space. The orbital period distribution of \cite{davisetal10-1}
predicts a long tail of systems with periods $>1$\,d, inconsistent
with the period distribution of the PCEBs known at the time of
publication (their Fig.\,10), even when taking into account the
detection pro\-ba\-bi\-li\-ties of \cite{rebassa-mansergasetal08-1}. Our more
detailed observed (Fig.\ref{g:porbs58}) and bias-corrected
(Fig.\,\ref{g:biases}) orbital period distributions are in better 
agreement with their predicted orbital period distribution limited 
to PCEBs with companion masses $\le0.5$\,\Msun, i.e. spectral types later than 
$\sim$M0 (see their Fig.\,10 middle panel). 

In summary, most published BPS models predict a substantial number of
PCEBs with orbital periods $>1$\,d, which is cu\-rren\-tly incompatible
with the orbital period distribution of the SDSS PCEBs. The observed
dearth of long-period PCEBs could be reconciled with a low value of
$\alpha_\mathrm{CE}$. \cite{zorotovicetal10-1} used a sub-set of 35 of
the PCEBs presented here, along with some other 25 non-SDSS PCEBs, to
constrain the value of $\alpha_\mathrm{CE}$ by reconstructing their
evolution. They found that there seems to be no dependence of
$\alpha_\mathrm{CE}$ on the mass of the secondary star, and that if
the value is universal only a small value of $\alpha_\mathrm{CE}$ can
explain the short orbital periods observed among the known
PCEBs. However, it is also clear that even the SDSS PCEB sample is
subject to selection effect, i.e. systems with com\-pa\-nion spectral
types earlier than $\sim$M0, i.e. masses $\ga0.5$\,\Msun\, are severely
underrepresented, which are expected to have longer orbital periods
than the currently dominant population of PCEBs with low-mass (mid
M-type) companions.

\section{Conclusions}
Nearly 2000 WDMS binaries have been spectroscopically identified
within the SDSS survey. A detailed characterisation and analysis of
this large and homogeneously selected sample promises to improve 
substantially our understanding of close binary evolution, in particular of
the common envelope phase. Using follow-up observations of 385 WDMS
binaries, combined with the SDSS subspectra for an additional 864
systems, we identified 191 PCEBs. We determine orbital
periods for 58 of these systems, increasing the total number of SDSS
PCEBs with known orbital periods to 79. We analyse the observational
biases inherent to the SDSS PCEB sample, and find that $21-24$\% of
all SDSS WDMS binaries with M dwarf companions must have 
undergone common envelope
evolution. The bias-corrected orbital period distribution of the SDSS
PCEBs shows a peak near $\sim10.3$\,h, and drops fairly steeply to both
shorter and longer periods, with no systems at $\Porb<1.9$\,h, and few
systems at $P>1$\,d. Comparing the properties of the SDSS PCEB population to
the results of published binary population, we conclude that the
observed dearth of long-period systems can be explained by low values
of $\alpha_\mathrm{CE}$. 
Observational selection effects against earlier-type companions 
can also play a role. Given a Roche lobe-filling giant or 
asymptotic giant (the progenitor of the white dwarf component) in a binary 
system, the more massive the companion star, the greater the orbital energy 
available to eject the common envelope, and therefore the longer the orbital 
period of the remnant PCEB, other things being equal.
Unlocking the diagnostic potential of the
SDSS WDMS binary population now requires the computation of a
state-of-the art binary population model that includes a realistic
simulation of the observational selection effects inherent to this
sample. 

\begin{acknowledgements}

Based on observations collected at:
the European Organisation for Astronomical Research in the Southern
Hemisphere, Chile (078.D-0719, 079.D-0531, 080.D-0407, 082.D-0507,
083.D-0862, 085.D-0974);
the Gemini Observatory, which is operated by the Association of
Universities for Research in Astronomy, Inc., under a cooperative
agreement with the NSF on behalf of the Gemini partnership: the
National Science Foundation (United States), the Science and
Technology Facilities Council (United Kingdom), the National Research
Council (Canada), CONICYT (Chile), the Australian Research Council
(Australia), Minist\'{e}rio da Ci\^{e}ncia e Tecnologia (Brazil) and
Ministerio de Ciencia, Tecnolog\'{i}a e Innovaci\'{o}n Productiva
(Argentina) (GS-2007B-Q-27, GS-2008A-Q-31, GS-2008B-Q-40,
GS-2009A-Q-56); 
the Magellan Telescopes Baade and Clay, and the Ir\'en\'e Dupont
Telescope located at Las Campanas Observatory, Chile;
the William Herschel Telescope, operated on the island of La Palma by
the Isaac Newton Group in the Spanish Observatorio del Roque de los
Muchachos of the Instituto de Astrof\'isica de Canarias;
the IAC80 operated on the island of Tenerife by the Instituto de
Astrof\'isica de Canarias in the Spanish Observatorio del Teide;
and at the Centro Astron\'omico Hispano Alem\'an (CAHA) at Calar Alto,
operated jointly by the Max-Planck Institut f\"ur Astronomie and the
Instituto de Astrof\'isica de Andaluc\'a.

Funding for the SDSS and SDSS-II has been provided by the Alfred
P. Sloan Foundation, the Participating Institutions, the National
Science Foundation, the U.S. Department of Energy, the National
Aeronautics and Space Administration, the Japanese Monbukagakusho, the
Max Planck Society, and the Higher Education Funding Council for
England. The SDSS Web Site is http://www.sdss.org/.
%The  SDSS   is  managed  by  the  Astrophysical
%Research   Consortium   for   the  Participating   Institutions.   The
%Participating Institutions are the American Museum of Natural History,
%Astrophysical  Institute Potsdam, University  of Basel,  University of
%Cambridge,  Case Western  Reserve University,  University  of Chicago,
%Drexel  University, Fermilab,  the Institute  for Advanced  Study, the
%Japan  Participation  Group,   Johns  Hopkins  University,  the  Joint
%Institute for  Nuclear Astrophysics, the Kavli  Institute for Particle
%Astrophysics and  Cosmology, the  Korean Scientist Group,  the Chinese
%Academy  of Sciences  (LAMOST),  Los Alamos  National Laboratory,  the
%Max-Planck-Institute  for Astronomy  (MPIA),  the Max-Planck-Institute
%for  Astrophysics  (MPA),  New  Mexico State  University,  Ohio  State
%University,  University  of   Pittsburgh,  University  of  Portsmouth,
%Princeton  University, the  United States  Naval Observatory,  and the
%University of Washington.

The authors would like to thank the referee, R.~Webbink, for carefully reading 
the manuscript and for providing useful remarks which helped improving the 
quality of the paper. 
We would like to thank T.~Forveille and B.~Miszalski for useful comments.
ANGM acknowledges support by the Centre National d'Etudes Spatial
(CNES, ref. 60015).  
BTG was supported by a rolling grant from STFC. 
ANGM and MRS acknowledge support by the Deutsches Zentrum f\"ur Luft und Raumfahrt (DLR) GmbH under contract No. FKZ 50 OR 0404. 
MRS was also supported by FONDECYT (grant 1100782), and the Center of Astrophysics at the Universidad de Valparaiso. 
ARM acknowledges financial support from FONDECYT (grant 3110049). 
This work was supported by the Deutsche Forschungsgemeinschaft (DFG) under contract numbers Schw 536/26-1, 29-1, 30-1, 31-1, 32-1, 33-1, and 34-1.
JS acknowledges financial support from STFC in the form of an Advanced Fellowship. 
\end{acknowledgements}

\bibliographystyle{aa}
\bibliography{aamnem99,references}
\Online
\begin{appendix}
\section{Optical light curves}
\label{sec:app2}
\begin{figure*}[t!]
\begin{center}
\includegraphics[angle=0,clip=,width=0.45\textwidth]{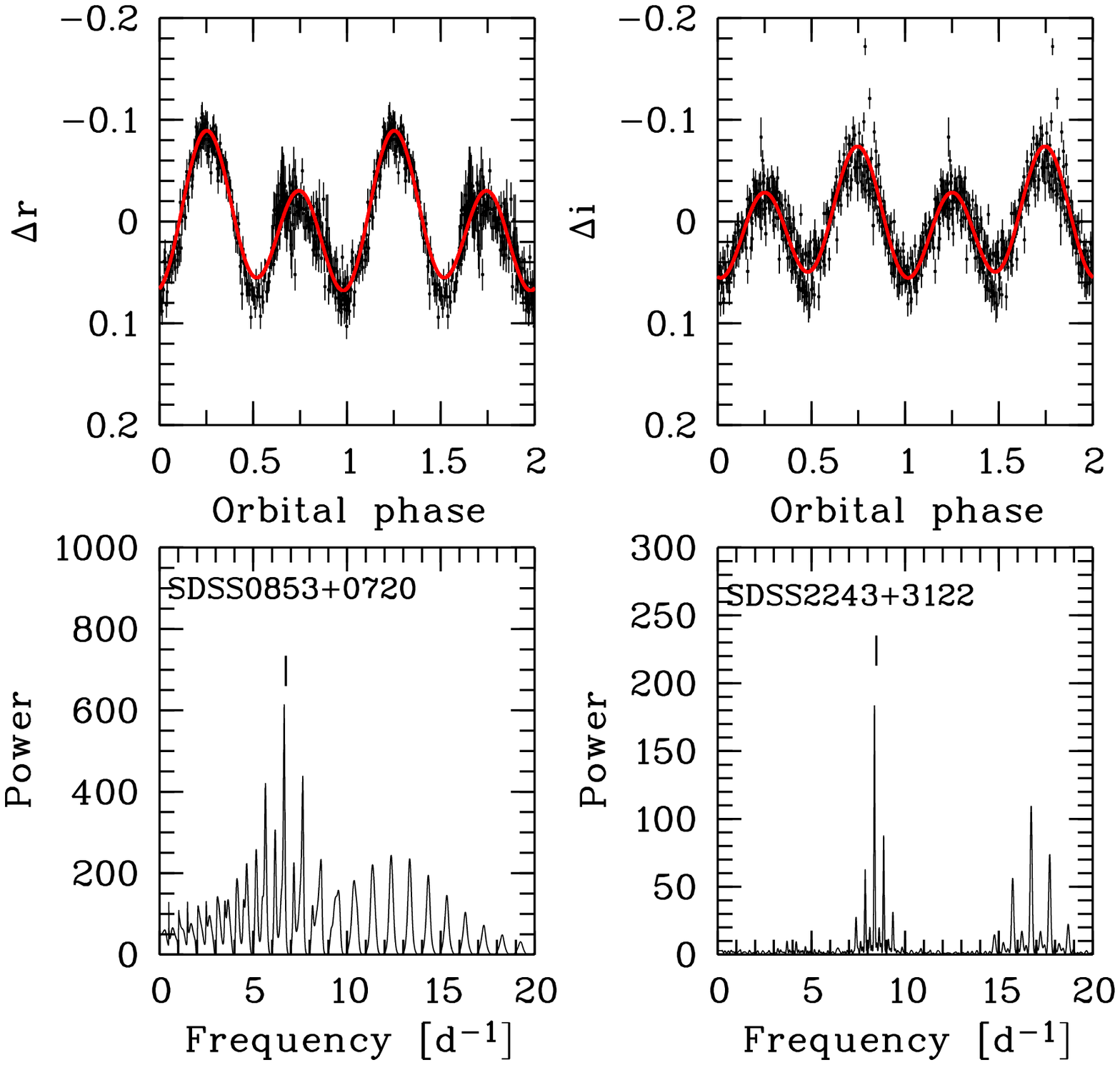}
\includegraphics[angle=0,clip=,width=0.45\textwidth]{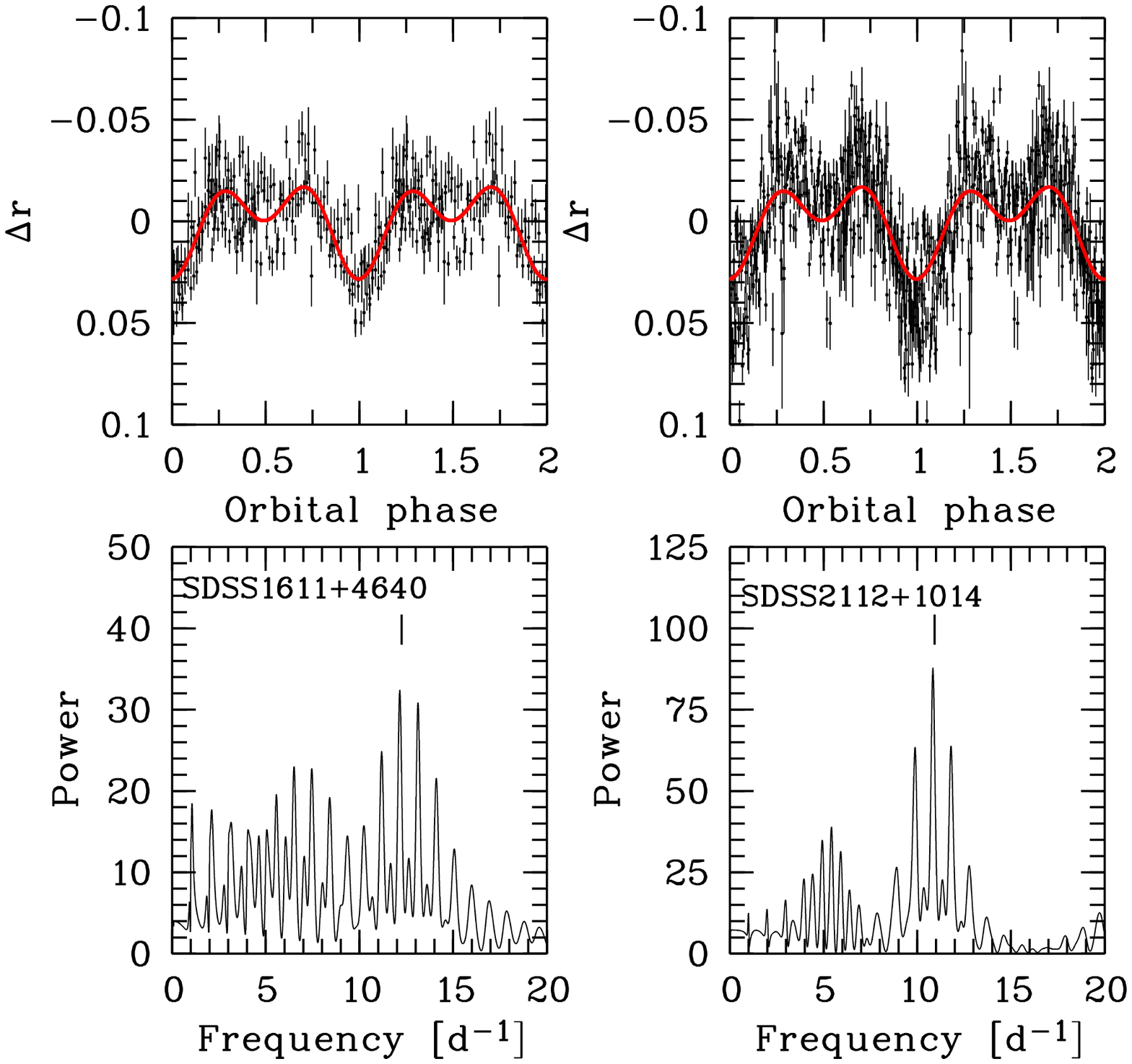}
\caption [Light  curve and radial velocity  curve for SDSS\,0853+0720,
SDSS\,2243+3122,                  SDSS\,1611+4640                  and
SDSS\,2112+1014]{\label{g:sdss0853_2243}\label{g:sdss1611_2112}Photometric
periodogram,   and  photometric   light   curves  for   SDSS0853+0720,
SDSS\,2243+3122, SDSS\,1611+4640 and SDSS\,2112+1014 phase folded with
the  orbital   period  corresponding  to  the  highest   peak  in  the
periodogram  (bottom  panels),  \Porb=$0.1503$\,d,  \Porb=$0.11954$\,d, 
\Porb=$0.0823$\,d and \Porb=$0.0923$\,d respectively. Double-sine
fits are shown in red.}
\end{center}
\end{figure*}
Photometric follow-up observations were carried out for a total of seven 
WDMS binaries where the available spectroscopy suggested a short orbital 
period. Telescope, filter and duration of the observations are listed in 
Table~\ref{t:log}. 
Four out of the seven observed systems showed ellipsoidal modulation:
SDSS\,0853+0720, SDSS\,1611+4640, SDSS\,2112+1014, and SDSS\,2243+3122. 
We determined their orbital periods by computing periodograms and
fitting sine curves to the phase folded light curves (see Fig.~\ref{g:sdss0853_2243}). 
We estimated $\Porb=3.6, 1.9, 2.2,$ and $2.8$ hours respectively. 
Spectroscopic follow-up observations were carried out for 
SDSS\,0853+0720, SDSS\,1611+4640, and SDSS\,2243+3122 covering a full cycle,
and confirming the measured periods. 
We could connect the spectroscopic and photometric runs for SDSS\,1611+4640, 
and SDSS\,2243+3122, without cycle count aliases, which allowed us to refine 
their orbital periods.

\begin{figure}[h!]
\begin{center}
\includegraphics[angle=-90,clip=,width=0.45\textwidth]{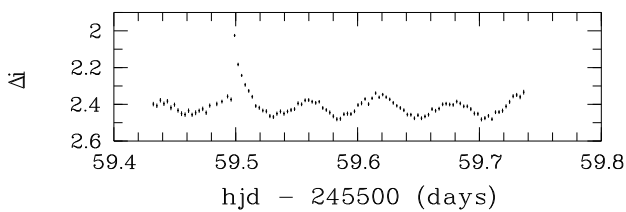}
\caption   [Flare   in   SDSS\,2243+2122]{\label{g:flare}   Differential
photometry in the \emph{i}  band obtained for SDSS\,2243+2122 the 15th
August  2009. A  flare  of  $\sim25$ minutes  length  with a  relative
enhancement of $\sim 0.4$ magnitudes was observed. }
\end{center}
\end{figure}

The light curves of SDSS\,0853+0720 and SDSS\,2243+3122 show both two 
uneven maxima at phases $0.25$ and $0.75$. 
For SDSS\,0853+0720  the first maximum is brighter than 
the second one, while for SDSS\,2243+3122 is the other way round. In 
both systems equal minima at phases $0$ and $0.5$ are observed and   
the maximum variation is almost $0.2$ magnitudes in the \emph{r} and the 
\emph{i} bands respectively. 
Ellipsoidal modulation and spots in one of the hemispheres of the 
secondary star could explain 
the uneven maxima (O'Connell effect, see \cite{liuandyang03-1}). 
Photometry carried out on the 15th of August 2009 for SDSS\,2243+3122 
revealed a  flare with $\sim25$ minutes length  in the decay, and
rising   by   $0.35$   magnitudes    in   the   \emph{i}   band   (see
Figure~\ref{g:flare}). This would  strengthen the argument of magnetic
activity  being  the  cause  of  the  observed disparity  in  the  
maxima  (see Figure~\ref{g:sdss0853_2243}). 

Although no radial velocity curve is available for SDSS\,2112+1014, 
the five available spectra show radial velocity variations of more than 300 \kms\, 
in less than 40 minutes, indicating a rather short orbital period, consistent with
the measured photometric orbital period of $2.2$ hours. 

\section{Notes on individual systems}
\textbf{SDSS\,0225+0054}, the  primary star was classified as  a DC by
\cite{kleinmanetal04-1}. Although  it is true that  no absorption lines
are visible in  the primary star we consider that  the signal to noise
of the residual white dwarf  after the subtraction of the best fitted
secondary template  is insufficient for giving any other classification
than white dwarf \citep[see][for    details    on     the     spectral
analysis]{rebassa-mansergasetal07-1}.   SDSS\,0225+0054  together with
\textbf{SDSS\,0853+0720} are WDMS binaries consisting of cold DC white
dwarf and  an $\mathrm{M4.5}$ and $\mathrm{M3}$  companion star, clear
examples of the old PCEBs predicted by \citet{schreiber+gaensicke03-1}
that would not have been found in any search for blue objects.

\noindent
\textbf{SDSS\,0307+3848}  (SDSS\,J030716.44+384822.8)  has  a red  star
nearby - within 5\arcsec  - (SDSS\,J030716.16+384821.4) and is slightly
brighter in the r band (16.70 versus 15.17). To be sure that there was
no contamination in our spectra by  this source we rotated the slit so
as to  take spectra simultaneously and compared  the radial velocities
with those  of this  system. No variations  in the radial  velocity of
this  visual   companion  were  observed.  The   optical  spectrum  of
SDSS\,0307+3848  is dominated by  the secondary  star and  it presents
strong  emission lines.  We measured  the radial  velocities  from the
\Ha\, line in  the SDSS-subspectra. The \Na\, and the  \Ha\, lines are in
phase and  the amplitudes were  consistent with each other  within the
errors.

\noindent
\textbf{SDSS\,1006+0044},         \textbf{SDSS\,1523+4604}         and
\textbf{SDSS\,1844+4120} have  very cold  WDs, $\Teff =  7\,800, 8\,400,
7\,600$\,K,   late secondary star spectral  types $\mathrm{M9}$,   
$\mathrm{M7}$,
$\mathrm{M6}$, and are among the  closest PCEBs from the sample, $\dwd
= 60, 76, 73$ pc  respectively. SDSS\,1844+4120 is the system with the
highest  systemic  velocity  and  it  has Ca\,II  H\&K  in  absorption, 
i.~e. the white dwarf is a DAZ. This photospheric absorption 
could be originated by some wind from the secondary star that is being accreted
by the WD (see e.g. LTT560 \citep{tappertetal07-1}).
Looking   at  the   SDSS   images  of   SDSS\,1523+4604,  aka   CSO749
\citep{sanduleak+pesch89-1}, the components of  this binary seem to be
slightly  separated, also  noted by  \cite{helleretal09-1},  and there
seems to be  a third red object towards  the North-East. We have  
measured  an orbital period  of $9.93$  hours, so  it could  be that  
SDSS\,1523+4604  is a triple system formed by a close binary and a 
wide companion.

\noindent
\textbf{SDSS\,1143+0009}, aka WD1140+004,  is an X-ray emitting source
\citep{aguerosetal09-1}    from    the    Rosat   All    Sky    Survey
\citep{vogesetal99-1}.     The   X-ray    to   optical    flux   ratio
($\log\mathrm{f_X/f_g}$)  is 0.22,  a value too high to be explained  
by the coronal  emission of  the $\mathrm{dMe}$
star \citep[see][for      typical      values]{maccacaroetal88-1,
motchetal98-1},  and the  SDSS spectrum  presents some  marginal \Ha\,
emission (although  this can  be partially masked  by the flux  of the
white dwarf). The high flux ratio could be explained if some wind from
the secondary star is being accreted  by the white dwarf. If so, some
Ca\,II  H\&K absorption would  then  be expected,  which  is not  visible 
in  the spectrum.  
%Metal lines in the WD photosphere would be a good sign.
A small fraction of isolated white dwarfs show hard (higher
than            0.5           keV)            X-ray           emission
\citep{o'dwyeretal03-1,chuetal04-1,bilikovaetal10-1}, nevertheless the
effective  temperature of  the white  dwarf in  SDSS\,1143+0009 ($\sim
17~000$\,K) is  too cold to explain it's hard  X-ray emission. 
\citet{briggsetal07-1} list other six PCEBs X-ray
emitters: BPM~71214,  RR~Cae,  UZ~Sex, EG~UMa, Feige~24 and  WD~1541-381 . 
We investigated  the X-ray fluxes  of these
sources      within     the      XMM-Newton     making      use     of
xcat-DB\footnote{http://xcatdb.u-strasbg.fr/}
\citep{micheletal04-1}. None  of them show an X-ray spectrum as hard as
that of SDSS\,1143+0009.  
It has  a  counterpart  the  Galaxy  Evolution  Explorer
(GALEX)  database \citep{martinetal05-1}, with  detection only  in the
FUV  flux,  $17.97  \pm  0.08$  mag,  consistent  with  the  effective
temperature of the white dwarf. We conclude that there is no 
straight-forward explanation of the hard X-ray emission of SDSS\,1143+0009. 

\noindent
\textbf{SDSS\,1352+0910} contains a hot WD,  $\Teff = 36~200$\,K and no
companion visible in SDSS spectrum. Nevertheless emission in the \Ha\,
line was observed in the SDSS spectrum, pointing to a binary nature of
the system,  and triggering further spectroscopy which  revealed it to
be a close binary. We investigated infrared  surveys to learn more about 
the secondary star. We found  a counterpart in  the UKIDSS
catalogue,  ULAS  J135228.14+091039.0.   For  details  on  the  UKIDSS
project, camera, photometric  system, calibration, pipeline processing
and   science  archive  see   \cite{lawrenceetal07-1,  casalietal07-1,
hewettetal06-1,  hodgkinetal09-1,hamblyetal08-1}. Magnitudes  found in
the 7th data release are Y $= 18.24 \pm 0.03$, J $= 18.04 \pm 0.03$, H
= $+17.68  \pm 0.04$, K =  $17.49 \pm 0.06$ mag,  yielding infrared
colours  J-H  = 0.36  and  H-K =  0.19  magnitudes.  We subtracted  the
infrared contribution of the white dwarf and, 
using     the      empirical     luminosity-mass     relation     from
\cite{delfosseetal00-1} we  estimated a mass of 0.175  \Msun\, for the
secondary  star. Using  the spectral  type mass  radius  relation from
\cite{rebassa-mansergasetal07-1}  this is  consistent with  a spectral
type of $\mathrm{M6.5}$.

\noindent
\textbf{SDSS\,1434+5335} (SDSS\,J143443.24+533521.2)  has a back\-ground
or  near\-by  red ob\-ject  (SDSS\,J143443.19+533525.0),  classified as  a
galaxy by  the SDSS. The background galaxy is  too faint,
and our  object is  sufficiently bright so  that integration  time was
relatively short  and a  possible contamination of our  spectra is negligible.

\noindent
\textbf{SDSS\,1436+5741} and \textbf{SDSS\,2156--0002} 
present a slight blue  excess in the SDSS spectra with respect to single 
dM stars. Spectroscopic observations carried out
for  these  two  systems  (see Table~\ref{t:log})  showed  significant
radial velocity  variation, confirming  their binary nature.  We explored
GALEX with the aim of learning about the primary stars and found that 
SDSS\,2156--0002 has GALEX\,J215614.5-000237  as counterpart. Only flux
in  the  NUV was  detected,  with a  magnitude  of  $21.99 \pm  0.34$,
consistent with a cold white dwarf.

\noindent
\textbf{SDSS\,1559+0356} has a hot white dwarf, $\Teff =48~212$\,K, and
no companion is  visible in the SDSS spectrum. It  has a strong \Ha\,
emission  line  from  which  we   were  able  to  measure  the  radial
velocities.  SDSS\,1559+0356   was  listed   as  a  CV   candidate  by
\cite{szkodyetal07-1}, but  there is no obvious sign  of mass transfer
in the spectrum. With its orbital period of $2.2$ hours this system is
located in the CV orbital period gap. If it would be a detached CV one
would expect a cooler white dwarf \citep{townsley+gaensicke09-1}.
We investigated infrared surveys to have more information for the 
secondary star, but no counterpart was found for the object.

\noindent
\textbf{SDSS\,1611+4640} has the shortest period among the PCEBs, with
$\Porb = 1.97$ hours,  and has a cold white dwarf, $\Teff \sim  10~300$\,K 
and a late secondary star spectral type, $\mathrm{M5}$.

\noindent
\textbf{SDSS\,1731+6233} has a  red nearby star, care was  taken so as
not to  contaminate our  spectra with any  contribution from  it.  
Unbeknownst to us at the start of our spectroscopic campaign, this system 
had been previously classified a probable binary by
\cite{pourbaixetal05-1} on the basis  of radial velocity variations of
$\sim  40$  \kms.

\noindent
\textbf{SDSS\,2149--0717} has  SDSS photometry but it was not  an SDSS
spec\-tros\-copic   target. It   was   identified   as    a   WDMS   by
\cite{southworthetal07-1} during some  pilot follow-up studies of WDMS
candidates in the  SDSS footprint at the UK  Schmidt telescope with the
6dF  spectrograph. The  spectral type  of the  secondary star  is less
certain than  for SDSS spectra since  the 6dF only  covers until 7~500\,\AA.

\noindent
\textbf{SDSS\,2208+1221} contains the hottest white dwarf in our sample, 
$\Teff \sim 100~000$\,K  and the earliest spectral  type secondary star, 
a $\mathrm{K7}$ star \citep{silvestrietal06-1}.

\noindent
\textbf{SDSS\,2243+3122} has a $\mathrm{M5}\pm1$ so that it could be a 
detached CV (dCV)
or hibernating CV, that has relaxed to its equilibrium radius.
The SDSS spectrum presents the Balmer lines in emission. 
Radial velocities from the \Ha\, emission line showed no 
difference in phasing from the \Na\, absorption doublet, i.e. both sets 
of lines must originate from the secondary which argues against a CV nature 
of the system. 

\end{appendix}
\end{document}

%% file: table1.tex
\begin{table*}
\begin{center}
\setlength{\tabcolsep}{1ex}
\caption[]{\label{t:log}Log of observations. %Object
%  name, \emph{ugriz} psf-magnitudes, telescope, and number of frames
%  taken for each system.
}
\begin{tabular}{ccccccrl}
\hline\hline 
\noalign{\smallskip}
Object                   & u     & g     & r     & i     & z     &
Spectroscopy: Telescope  & \# of spec. \\
\noalign{\smallskip}
\hline
\noalign{\smallskip}
\object{SDSS\,J015225.38\,--\,005808.5}  & 18.36 & 17.95 & 17.70 & 16.98 & 16.22 & Baade,NTT,WHT      & 26 \\
\object{SDSS\,J022503.02\,+\,005456.2}  & 21.37 & 20.55 & 19.65 & 18.35 & 17.47 & Baade              & 16 \\
\object{SDSS\,J023804.39\,--\,000545.7}  & 19.81 & 19.23 & 18.68 & 17.70 & 16.97 & Baade,WHT          & 16 \\
\object{SDSS\,J030138.24\,+\,050218.9}  & 19.11 & 18.39 & 18.05 & 17.27 & 16.61 & VLT,NTT          & 21 \\
\object{SDSS\,J030544.41\,--\,074941.2}  & 18.68 & 18.33 & 18.12 & 17.30 & 16.70 & Baade,Clay,NTT,CA3.5,WHT & 52 \\
\object{SDSS\,J030716.44\,+\,384822.8}  & 20.64 & 19.04 & 17.86 & 16.70 & 16.06 & CA3.5             & 24 \\
\object{SDSS\,J032038.72\,--\,063822.9}  & 19.77 & 19.28 & 19.16 & 18.52 & 17.89 & Baade,NTT,VLT,WHT  & 25 \\
\object{SDSS\,J080736.96\,+\,072412.0}  & 19.44 & 19.13 & 18.84 & 18.05 & 17.46 & VLT,Baade,Clay   & 27 \\ 
\object{SDSS\,J083354.84\,+\,070240.1}  & 19.15 & 18.73 & 18.75 & 18.15 & 17.60 & VLT,Clay           & 19 \\
\object{SDSS\,J085336.03\,+\,072033.5}  & 19.99 & 19.21 & 18.53 & 17.47 & 16.70 & VLT,Baade          & 14 \\
\object{SDSS\,J092452.39\,+\,002449.0}  & 18.17 & 18.03 & 18.22 & 17.87 & 17.41 & NTT,VLT          & 17 \\
\object{SDSS\,J094913.37\,+\,032254.5}  & 19.30 & 18.99 & 18.95 & 18.31 & 17.64 & VLT,Clay          & 22 \\
\object{SDSS\,J100609.18\,+\,004417.0}  & 18.32 & 17.85 & 17.70 & 17.36 & 16.68 & Gemini-S,Clay          & 14 \\
\object{SDSS\,J110517.60\,+\,385125.7}  & 19.05 & 18.44 & 18.08 & 17.32 & 16.74 & CA3.5             & 16 \\
\object{SDSS\,J114312.57\,+\,000926.5}  & 18.43 & 18.15 & 18.17 & 17.60 & 17.03 & Gemini-N,NTT          & 13 \\
\object{SDSS\,J123139.80\,--\,031000.3}  & 19.41 & 18.67 & 18.19 & 17.09 & 16.26 & NTT,VLT          & 18 \\
\object{SDSS\,J130012.49\,+\,190857.4}  & 19.85 & 19.22 & 18.68 & 17.66 & 16.89 & CA3.5             & 16 \\
\object{SDSS\,J131334.74\,+\,023750.7}  & 19.12 & 18.67 & 18.53 & 17.73 & 17.05 & VLT,Baade,NTT,WHT  & 37 \\
\object{SDSS\,J131632.04\,--\,003758.0}  & 19.76 & 18.75 & 18.25 & 18.06 & 17.90 & CA3.5             & 10 \\
\object{SDSS\,J134841.61\,+\,183410.5}  & 17.70 & 17.31 & 17.19 & 16.47 & 15.88 & CA3.5             & 24 \\
\object{SDSS\,J135228.14\,+\,091039.0}  & 17.85 & 18.13 & 18.56 & 18.81 & 18.83 & Gemini-S,Clay          & 17 \\
\object{SDSS\,J141134.70\,+\,102839.7}  & 18.74 & 18.90 & 19.14 & 19.09 & 18.91 & Gemini-S,Clay          & 18 \\
\object{SDSS\,J142951.19\,+\,575949.0}  & 20.06 & 19.43 & 18.75 & 17.76 & 17.07 & CA3.5,WHT         & 21 \\
\object{SDSS\,J143443.24\,+\,533521.2}  & 15.96 & 15.97 & 16.26 & 16.19 & 15.92 & WHT              & 31 \\
\object{SDSS\,J143642.01\,+\,574146.3}  & 20.87 & 19.73 & 18.60 & 17.38 & 16.69 & CA3.5,WHT         & 24 \\
\object{SDSS\,J143746.69\,+\,573706.0}  & 18.96 & 18.52 & 18.11 & 17.06 & 16.35 & CA3.5             & 10 \\
\object{SDSS\,J143947.62\,--\,010606.7}  & 16.14 & 16.55 & 16.76 & 16.72 & 16.61 & VLT,Baade,Clay,CA3.5,WHT & 42 \\
\object{SDSS\,J150657.58\,--\,012021.7}  & 19.61 & 19.09 & 18.94 & 18.26 & 17.67 & Baade,Clay,NTT,WHT,VLT  & 17 \\
\object{SDSS\,J151921.72\,+\,353625.8}  & 16.83 & 16.69 & 17.02 & 17.01 & 16.64 & WHT              & 22 \\
\object{SDSS\,J152359.22\,+\,460448.9}  & 18.23 & 17.92 & 17.76 & 17.18 & 16.40 & WHT              & 13 \\
\object{SDSS\,J152425.21\,+\,504009.7}  & 17.44 & 17.31 & 17.27 & 16.60 & 16.06 & WHT,CA3.5,CA2.2    & 47 \\
\object{SDSS\,J152821.45\,+\,344314.9}  & 18.45 & 18.43 & 18.46 & 17.86 & 17.28 & WHT,CA3.5         & 19 \\
\object{SDSS\,J155808.49\,+\,264225.7}  & 19.82 & 19.18 & 18.47 & 17.31 & 16.56 & CA3.5             & 29 \\
\object{SDSS\,J155904.62\,+\,035623.4}  & 18.23 & 18.40 & 18.58 & 18.68 & 18.68 & Gemini-S,Clay          & 18 \\
\object{SDSS\,J160821.47\,+\,085149.9}  & 19.12 & 18.70 & 18.65 & 18.16 & 17.53 & Gemini-S,Clay          & 17 \\
\object{SDSS\,J161145.88\,+\,010327.8}  & 19.17 & 18.64 & 18.51 & 17.94 & 17.28 & Gemini-N,CA3.5         & 13 \\
\object{SDSS\,J161113.13\,+\,464044.2}  & 18.28 & 17.72 & 17.84 & 17.48 & 17.02 & WHT              & 16 \\
\object{SDSS\,J162354.45\,+\,630640.4}  & 19.84 & 19.12 & 18.48 & 17.39 & 16.71 & CA3.5,WHT         & 29 \\
\object{SDSS\,J162552.91\,+\,640024.9}  & 19.51 & 19.09 & 18.91 & 18.31 & 17.56 & CA3.5,WHT         & 12 \\
\object{SDSS\,J164615.60\,+\,422349.2}  & 18.73 & 18.41 & 18.54 & 17.95 & 17.22 & WHT              & 20 \\
\object{SDSS\,J170509.65\,+\,210904.4}  & 17.77 & 17.77 & 18.00 & 17.60 & 17.06 & WHT,NTT          & 35 \\
\object{SDSS\,J171810.15\,+\,610114.0}  & 18.28 & 18.00 & 18.20 & 17.78 & 17.19 & WHT              & 23 \\
\object{SDSS\,J173101.49\,+\,623315.9}  & 18.58 & 18.27 & 18.31 & 17.74 & 17.10 & WHT              & 14 \\
\object{SDSS\,J184412.58\,+\,412029.4}  & 17.26 & 16.71 & 16.43 & 15.69 & 14.93 & CA3.5,WHT         & 21 \\
\object{SDSS\,J204541.90\,--\,050925.7}  & 19.04 & 19.04 & 19.06 & 18.51 & 17.99 & VLT,Baade,Clay,NTT,WHT  & 34 \\
\object{SDSS\,J211205.31\,+\,101427.9}  & 18.68 & 18.38 & 18.42 & 18.05 & 17.35 & Gemini-S,WHT          &  5 \\
\object{SDSS\,J211428.41\,--\,010357.2}  & 18.71 & 18.64 & 18.47 & 17.83 & 17.24 & Clay,VLT          & 26 \\
\object{SDSS\,J212051.92\,--\,005827.3}  & 18.09 & 17.75 & 17.73 & 17.11 & 16.48 & Clay,WHT          & 35 \\
\object{SDSS\,J212320.74\,+\,002455.5}  & 19.73 & 19.24 & 19.35 & 19.01 & 18.39 & Clay,VLT          & 13 \\
\object{SDSS\,J213218.11\,+\,003158.8}  & 18.40 & 18.03 & 18.16 & 17.89 & 17.49 & Clay,WHT          & 17 \\
\object{SDSS\,J214952.27\,--\,071756.9}  & 17.50 & 16.74 & 16.04 & 15.04 & 14.41 & WHT              & 21 \\
\object{SDSS\,J215614.57\,--\,000237.4}  & 20.77 & 20.14 & 19.55 & 18.28 & 17.31 & Baade,NTT,VLT,WHT  & 41 \\
\object{SDSS\,J220848.99\,+\,122144.7}  & 17.76 & 17.72 & 17.24 & 16.82 & 16.49 & CA3.5,Clay         & 19 \\
\object{SDSS\,J221616.59\,+\,010205.6}  & 18.37 & 17.85 & 17.99 & 17.89 & 17.45 & Gemini-S,WHT          & 14 \\
\object{SDSS\,J224038.37\,--\,093541.4}  & 18.01 & 17.54 & 17.55 & 17.14 & 16.54 & NTT,WHT          & 50 \\
\object{SDSS\,J224307.59\,+\,312239.1}  & 20.76 & 19.59 & 18.69 & 17.28 & 16.26 & CA3.5             & 12 \\
\object{SDSS\,J231105.66\,+\,220208.6}  & 19.61 & 19.07 & 18.38 & 17.45 & 16.80 & CA3.5             & 29 \\
\object{SDSS\,J231825.23\,--\,093539.1}  & 18.33 & 18.31 & 18.49 & 18.13 & 17.71 & Baade,NTT,WHT      & 39 \\
%\hline
%\hline
\noalign{\smallskip}
Object                    & u     & g     & r     & i     & z     &
Photometry: Telescope   & duration, filter  \\
\noalign{\smallskip}
\object{SDSS\,J023804.39\,--\,000545.7}  & 19.81 & 19.23 & 18.68 & 17.70 & 16.97 & Ca2.2       &   1.2 h, R        \\
\object{SDSS\,J085336.03\,+\,072033.5}  & 19.99 & 19.21 & 18.53 & 17.47 & 16.70 & DuPont     &   3.7 h, r        \\
\object{SDSS\,J161113.13\,+\,464044.2}  & 18.28 & 17.72 & 17.84 & 17.48 & 17.02 & Ca2.2       &     5 h, R        \\
\object{SDSS\,J211205.31\,+\,101427.9}  & 18.68 & 18.38 & 18.42 & 18.05 & 17.35 & IAC80,Ca2.2 &  19.3 h, I , R    \\
\object{SDSS\,J213218.11\,+\,003158.8}  & 18.40 & 18.03 & 18.16 & 17.89 & 17.49 & Ca2.2       &   4.5 h, R        \\
\object{SDSS\,J224038.37\,--\,093541.4}  & 18.01 & 17.54 & 17.55 & 17.14 & 16.54 & Ca2.2       &   3.5 h, R        \\
\object{SDSS\,J224307.59\,+\,312239.1}  & 20.76 & 19.59 & 18.69 & 17.28 & 16.26 & IAC80      &  26.47 h, i       \\
\noalign{\smallskip}\hline 
\end{tabular}  
\end{center}
\end{table*}

%% file: table2.tex
\setlength{\tabcolsep}{0ex}
\begin{table}
\begin{center}
\caption[Resolutions]{\label{t:res}Observational technical
  details. }
\fontsize{7}{7}\selectfont
\begin{tabular}{lccccc}
\hline\hline 
\noalign{\smallskip}
\multicolumn{6}{c}{Spectroscopy} \\
Tel./Inst. & Slit      & Spect.           & Spect.     & $<\sigma_{RV}>$ & \# of \\
           & width[\arcsec] & range[\AA]  & res.[\AA]& [\kms]          & RVs   \\
\noalign{\smallskip}
\hline                                 				      	 
\noalign{\smallskip}
VLT/FORS2      & 1      & 7830-9570  & 2.2         & 7.1             & 334  \\
Gemini-N/GMOS  & 1      & 7770-8470  & 3.5         & 7.1             & 124  \\
Gemini-S/GMOS  & 1      & 7770-8470  & 4.1         & 8.6             &  26  \\
Clay$\ddag$/LDSS3  & 0.75   & 5800-9980  & 4.8         & 10.7            & 302 \\
Baade$\ddag$/IMACS & 0.75   & 7700-8560  & 1.5         & 11.0            & 187  \\
WHT/ISIS       & 1      & 7600-9000  & 1.6         & 9.2             & 524  \\	
NTT/EMMI       & 1      & 7770-8830  & 2.8         & 11.4            & 284  \\
NTT/EFOSC2     & 1      & 7000-8600  & 6.5         & 19.5            & 94   \\
CA3.5/TWIN     & 1.5    & 7500-8500  & 1.6         & 11.6            & 443  \\
CA3.5/MOSCA    & 1.5    & 5400-9000  & 7           & 17.0            & 12   \\
SDSS           & 3$^\dag$ & 3900-9100  & 4           & 19.1            & 5168 \\
\hline                                  				      	 
\noalign{\smallskip}
\multicolumn{6}{c}{Photometry} \\
Tel./Inst.     && Resolution\,[\arcsec/pix.]     && \# of images \\
\noalign{\smallskip}
\hline                            				      	 
\noalign{\smallskip}
IAC80/CAMELOT  && 0.6    && 397 \\
DuPont/CCD     && 0.259  && 337 \\
CA2.2/CAFOS    && 0.53   && 1083    \\
\noalign{\smallskip}\hline 
\end{tabular}  
\end{center}
\textbf{Notes.}%Listed are teles\-cope/instrument, slit width, spectral
  %coverage, spectral resolution, mean error of the radial velocities,
  %and the number of radial velocity measurements for the spectroscopy,
  %and spatial resolution and number of images for the
  %photometry. 
  $^\dag$ SDSS uses a fibre-fed spectrograph. $^\ddag$~The two Magellan 
  6.5\,m telescopes.
\end{table}
\normalsize

%% file: table3.tex
\begin{table}
\begin{center}
\setlength{\tabcolsep}{0.4ex}
\caption[Rvs of identifications]{\label{t:rvs_iden} 
Radial velocities used for the identification of the binaries.} 
\fontsize{8}{8}\selectfont
\begin{tabular}{ccccr}
\hline\hline 
\noalign{\smallskip}
Name  & RV  & err  & hjd  & Teles. \\
SDSSJ & [\kms] & [\kms] & [days] & \\  
\noalign{\smallskip}
\hline
\noalign{\smallskip}
  $002749.99-001023.3$ &     13.2 &     15.3 &    2452262.628080 &           SDSS  \\
  $002749.99-001023.3$ &     11.0 &     17.3 &    2452262.640324 &           SDSS  \\
  $002749.99-001023.3$ &      7.0 &     17.0 &    2452262.652580 &           SDSS  \\
  $002749.99-001023.3$ &      8.4 &     10.5 &    2454347.439589 &           WHT  \\
  $002749.99-001023.3$ &     15.8 &      6.7 &    2454348.551886 &           WHT  \\
  $002801.68+002137.7$ &    -25.6 &     22.4 &    2452262.628105 &            SDSS  \\
  $002801.68+002137.7$ &      1.8 &     21.0 &    2452262.640349 &            SDSS  \\
  $002801.68+002137.7$ &      0.0 &     33.6 &    2452262.652605 &            SDSS  \\
  $002801.68+002137.7$ &     -4.8 &      5.0 &    2454381.694818 &            VLT  \\
  $002801.68+002137.7$ &     -3.3 &      5.0 &    2454382.804165 &            VLT  \\
  $002950.16+003225.8$ &     18.3 &     21.0 &    2452262.628153 &            SDSS  \\
  $002950.16+003225.8$ &      6.2 &     17.8 &    2452262.640397 &            SDSS  \\
  $002950.16+003225.8$ &     31.5 &     27.4 &    2452262.652652 &            SDSS  \\
  $002921.27-002032.8$ &    -62.6 &     28.0 &    2452262.628107 &            SDSS  \\
  $002921.27-002032.8$ &    -25.6 &     17.0 &    2452262.640351 &            SDSS  \\
  $002921.27-002032.8$ &    -22.3 &     18.7 &    2452262.652607 &            SDSS  \\
  $003221.86+073934.4$ &   -326.6 &      8.3 &    2454681.908960 &            Gemini-S  \\
  $003221.86+073934.4$ &    -27.9 &      5.9 &    2454687.871683 &            Gemini-S  \\
  $003336.49+004151.3$ &     67.8 &     17.2 &    2452930.850863 &            SDSS  \\
  $003336.49+004151.3$ &     90.9 &     16.4 &    2452932.782491 &            SDSS  \\
\noalign{\smallskip}\hline 
\end{tabular}
\end{center}
\end{table}
\normalsize

%% file: table4.tex
\begin{table}
\begin{center}
\setlength{\tabcolsep}{1ex}
\caption[Statistics observations]{\label{t:stats} 
%  Name of the system,
%  total number of spectra used for identification, number of spectra
%  from own spectroscopic observations, time span between first and
%  last observation and probability of measuring large radial velocity
%  variations.
Identification statistics.}
\begin{tabular}{ccccccrl}
\hline\hline 
\noalign{\smallskip}
Name  & N  & N$_{\mathrm{obs}}$ & $\Delta t$ [days] & Prob. \\
%      &    &                    &          &       \\
\noalign{\smallskip}
\hline
\noalign{\smallskip}
SDSSJ\,$000504.91+243409.6$ &   10 &    3 &       294.821 &   0.74189  \\ 
SDSSJ\,$000531.09-054343.2$ &    6 &    1 &       427.720 &   1.00000  \\
SDSSJ\,$000935.50+243251.2$ &   16 &    9 &       377.808 &   0.97822  \\
SDSSJ\,$001749.24-000955.3$ &    2 &    2 &         3.935 &   1.00000  \\
SDSSJ\,$002620.41+144409.5$ &    3 &    0 &         0.029 &   0.32658  \\
SDSSJ\,$003221.86+073934.4$ &    2 &    2 &         5.963 &   1.00000  \\
SDSSJ\,$004020.07+154156.3$ &    4 &    0 &        66.847 &   0.99943  \\
SDSSJ\,$005008.21-000359.0$ &    4 &    0 &        24.004 &   0.07446  \\
SDSSJ\,$005245.11-005337.2$ &    6 &    5 &      2127.066 &   0.99999  \\
\noalign{\smallskip}\hline 
\end{tabular}
\end{center}
\textbf{Notes.}
  Name of the system,  
  total number of spectra used for identification, number of spectra
  from own spectroscopic observations, time span between first and
  last observation and probability of measuring large radial velocity
  variations.
\end{table}

%% file: table5.tex
\addtolength{\tabcolsep}{0.65ex}
\begin{landscape}
\begin{table}
\begin{center}
\caption[Orbital period of PCEBs]{
%Orbital period, radial velocity amplitude, systemic velocity, type, and 
Stellar and binary parameters derived for the $58$ PCEBs presented in this work.}% Errors in the last digits are given in parenthesis. 
\label{t:allparam}       
\fontsize{7}{7}\selectfont
\begin{tabular}{lccccccccccccccccc} %cccccccccccccc}
\hline\hline 
\noalign{\smallskip}
Object &    \Porb        &  \Ksec            & $\gamma_\mathrm{sec}$ &  Type &    \Teff  & \logg      & \Mwd   & \dwd       & Sp2  & \dsec     &    \Msec   &   incl        &  $\mathrm{K}_{wd}$  & a        & \Rsec   & \Rsec/R$_{lob}$ \\
 SDSS  &    [h]          &  [km\,s$^{-1}$]   & [km\,s$^{-1}$]        &       &    [K]    &            & [\Msun]   &  [pc]      & [dM] & [pc]      &    [\Msun] &   [deg]       &  [km\,s$^{-1}$]     & [\Rsun]  &[\Rsun]  &                 \\
\noalign{\smallskip}
\hline\noalign{\smallskip}    
\noalign{\smallskip}
%# systems,Porb_h,ePh,Ksec/1e5,eKsec,gamma,egamma,type,Teff,err_Teff,logg,err_logg,Mwd,err_Mwd,dwd,err_dwd,Sp2,d2,err_d2,Msec,err_Msec1,sin_i_up,-,sin_i_low,asin(sin_i_up)*180/!pi,-,asin(sin_i_low)*180/!pi,Kwd,eKwd,aa/Rsun,(aa_max/Rsun-aa_min/Rsun)/2.,aa_min/Rsun,aa_max/Rsun,R_lob/Rsun,R_lob_min/Rsun,R_lob_max/Rsun,R,errR,R/(R_lob/Rsun),((R+errR)/(R_lob_min/Rsun)-(R-errR)/(R_lob_max/Rsun))/2.
$0152-0058$    & 2.1519468(99) &  136.1(3.1) &    1.5(2.1) &  DA/M&   8773(25)   &   8.19(09) &   0.72(06) &   98(6)   &   6 &  105(49)  &   0.20(08) &    21  -    22&    37(15) & 0.79(04) & 0.20(09) & 0.88(75) \\
$0225+0054$     &     21.86(16) &   88.5(5.1) &    1.3(5.0) &  WD/M&      -       &      -     &      -     &    -      & 4.4 &  408(147) &   0.32(09) &        -      &     -     & -        & 0.30(10) & -        \\
$0238-0005$    &    5.0801(48) &  254.6(2.9) &  -13.7(2.5) &  DA/M&  20566(1537) &   7.72(28) &   0.48(15) &  768(134) &   3 &  529(104) &   0.38(07) &   $\sim$90    &   201(71) & 1.42(12) & 0.39(08) & 0.57(31) \\
$0301+0502$     &  12.93874(47) &  176.8(2.7) &   11.1(2.1) &  DA/M&  11045(133)  &   8.41(09) &   0.86(05) &  160(11)  &   4 &  324(95)  &   0.32(09) &    58  -    65&    65(18) & 2.82(12) & 0.33(10) & 0.35(19) \\
$0305-0749$    &   48.4612(17) &   67.2(2.3) &   43.6(2.0) &  DA/M&  17707(410)  &   7.81(09) &   0.52(05) &  365(22)  &   4 &  331(97)  &   0.32(09) &    40  -    46&    41(12) & 6.35(35) & 0.33(10) & 0.12(07) \\
$0307+3848$     &   10.3266(23) &  145.1(1.7) &   -9.8(1.2) & DA:/M&      -       &      -     &      -     &   -       &   3 &  373(73)  &   0.38(07) &        -      &     -     & -        & 0.39(08) & -        \\
$0320-0638$    &    3.3751(18) &  287.3(3.5) &   -3.8(2.8) &  DA/M&  11173(361)  &   8.30(25) &   0.79(16) &  248(44)  &   5 &  423(217) &   0.25(12) &    56  -    90&    92(47) & 1.08(10) & 0.26(13) & 0.77(86) \\
$0807+0724$     &  11.45451(17) &  119.9(5.1) &    2.7(3.0) &  DA/M&  18542(760)  &   7.98(16) &   0.61(10) &  534(56)  &   3 &  624(123) &   0.38(07) &    41  -    52&    74(18) & 2.51(14) & 0.39(08) & 0.36(16) \\
$0833+0702$     &     7.344(21) &  173.9(5.8) &   88.0(4.2) &  DA/M&  15246(560)  &   7.87(12) &   0.54(07) &  386(30)  &   4 &  540(159) &   0.32(09) &    59  -    89&   102(31) & 1.77(11) & 0.33(10) & 0.44(27) \\
$0853+0720$     &    3.6091(25) &  226.2(4.7) &  -10.4(3.5) &  DC/M&      -       &      -     &      -     &   -       &   3 &  573(113) &   0.38(07) &        -      &     -     & -        & 0.39(08) & -        \\
$0924+0024$     &      57.7( 2) &   96.0(11.0)&    1.7(6.3) &  DA/M&  19193(335)  &   7.81(07) &   0.52(03) &  335(15)  &   4 &  514(151) &   0.32(09) &    84  -    90&    58(17) & 7.12(35) & 0.33(10) & 0.11(06) \\
$0949+0322$     &     9.491(15) &  158.2(4.0) &   92.3(2.8) &  DA/M&  18542(737)  &   7.79(15) &   0.51(08) &  549(52)  &   4 &  539(159) &   0.32(09) &    60  - 90 &    98(31) & 2.07(14) & 0.33(10) & 0.36(23) \\
$1006+0044$     &    6.7313(72) &  210.7(5.6) &    9.7(3.8) &  DA/M&   7819(61)   &   8.52(14) &   0.93(08) &   60(6)   &   9 &   38(1)   &   0.12(01) &    43  -    48&    26(3)  & 1.71(05) & 0.11(09) & 0.35(32) \\
$1105+3851$     &     8.274(11) &  152.6(3.7) &   -9.4(2.9) &  DA/M&  10548(44)   &   8.17(06) &   0.71(04) &  206(9)   &   3 &  462(91)  &   0.38(07) &    46  -    50&    81(15) & 2.10(07) & 0.39(08) & 0.46(17) \\
$1143+0009$     &     9.273(30) &  180.9(7.9) &   14.0(6.8) &  DA/M&  16910(349)  &   7.97(08) &   0.60(05) &  296(16)  &   4 &  415(122) &   0.32(09) &    69  -    90&    96(28) & 2.15(11) & 0.33(10) & 0.38(22) \\
$1231-0310$    &    5.8493(94) &  150.9(5.3) &   36.9(4.3) &  DA/M&  10073(200)  &   8.51(25) &   0.93(15) &  162(30)  &   4 &  306(90)  &   0.32(09) &    30  -    37&    51(16) & 1.61(10) & 0.33(10) & 0.63(40) \\
$1300+1908$     &     7.391(13) &  138.4(4.4) &  -14.4(3.2) &  DA/M&   8673(121)  &   8.81(19) &   1.09(10) &  124(21)  &   4 &  378(111) &   0.32(09) &    28  -    32&    40(12) & 2.09(10) & 0.33(10) & 0.53(31) \\
$1313+0237$     &  22.31916(45) &   83.8(3.1) &  -11.0(2.3) &  DA/M&  17912(469)  &   8.09(10) &   0.67(06) &  373(25)  &   4 &  408(120) &   0.32(09) &    32  -    36&    39(11) & 3.99(21) & 0.33(10) & 0.22(12) \\
$1316-0037$    &      9.66( 2) &  195.5(3.9) &  -12.9(3.2) &  DC/M&      -       &      -     &      -     &   -       &   3 &  394(78)  &   0.38(07) &        -      &     -     & -        & 0.39(08) & -        \\
$1348+1834$     &     5.962( 1) &  222.0(3.8) &  -20.7(2.9) &  DA/M&  15071(167)  &   7.96(03) &   0.59(02) &  180(3)   &   4 &  236(69)  &   0.32(09) &   $\sim$90    &   120(34) & 1.44(06) & 0.33(10) & 0.56(29) \\
$1352+0910$$^*$ &     8.295(26) &   57.1(3.4) &   47.0(3.0) &  DA/M&  36154(711)  &   7.49(11) &   0.45(04) &  856(74)  &  -1$^{**}$ &    -      &   -        &        -      &     -     & -        & -        & -        \\
$1411+1028$$^*$ &    4.0160(50) &  168.3(4.3) &   42.6(4.3) &  DA/M&  30419(701)  &   7.78(16) &   0.54(08) &  829(91)  &   3 & 1491(294) &   0.38(07) &    44  -    57&   118(27) & 1.25(07) & 0.39(08) & 0.68(29) \\
$1429+5759$     &  13.08486(22) &  140.5(3.9) &  -20.4(2.4) &  DA/M&  16149(109)  &   8.75(25) &   1.07(13) &  329(70)  &   3 &  632(124) &   0.38(07) &    37  -    43&    49(11) & 3.19(15) & 0.39(08) & 0.38(16) \\
$1434+5335$     &  104.5622(73) &   80.0(1.6) &  -40.4(1.2) &  DA/M&  21785(209)  &   7.74(04) &   0.49(02) &  153(4)   &   4 &  286(84)  &   0.32(09) &   $\sim$90    &    52(14) & 0.51(46) & 0.33(10) & 0.07(04) \\
$1436+5741$     &  20.73784(71) &  120.4(2.3) &  -15.7(1.9) &  WD/M&      -       &      -     &      -     &   -       &   3 &  511(101) &   0.38(07) &        -      &     -     & -        & 0.39(08) & -        \\
$1437+5737$     &  20.13369(38) &   65.2(6.2) &   -2.7(4.7) &  DA/M&  17707(675)  &   8.26(15) &   0.78(09) &  319(35)  &   4 &  323(95)  &   0.32(09) &    21  -    24&    26(8)  & 3.88(22) & 0.33(10) & 0.24(14) \\
$1439-0106$    &   36.5426(14) &   37.7(2.9) &    9.7(2.1) & DA/K:& 84455(5642)  &   8.11(19) &   0.81(09) &  383(65)  &   0 &  826(194) &   0.47(00) &    16  -    18&    21(2)  & 6.04(15) & 0.49(00) & 0.19(02) \\
$1506-0120$    &  25.22355(39) &  134.6(4.5) &  -26.3(3.0) &  DA/M&  15601(753)  &   7.68(19) &   0.45(09) &  498(56)  &   4 &  526(155) &   0.32(09) &   $\sim$90    &    95(33) & 3.99(32) & 0.33(10) & 0.18(12) \\
$1519+3536$     &     32.81(32) &   65.2(2.3) &   -0.8(1.8) &  DA/M&  19416(246)  &   7.90(05) &   0.57(03) &  165(6)   &   6 &  153(71)  &   0.20(08) &    29  -    30&    22(9)  & 4.70(22) & 0.20(09) & 0.13(11) \\
$1523+4604$     &    9.9317(51) &  202.5(1.9) &   -5.0(1.4) &  DA/M&   8378(49)   &   8.28(10) &   0.78(06) &   76(5)   &   7 &   73(40)  &   0.15(07) &    56  -    63&    38(18) & 2.15(10) & 0.15(08) & 0.29(30) \\
$1524+5040$     & 14.151167(36) &  167.5(2.3) &   -0.8(1.5) &  DA/M&  19416(292)  &   8.13(06) &   0.70(04) &  212(9)   &   3 &  336(66)  &   0.38(07) &    74  -    90&    90(17) & 3.04(10) & 0.39(08) & 0.32(12) \\
$1528+3443$     &     33.87(23) &  112.5(2.8) &  -79.9(2.1) &  DA/M&  26801(1022) &   7.85(14) &   0.56(07) &  554(51)  &   4 &  449(132) &   0.32(09) &    64  -    90&    64(19) & 5.03(31) & 0.33(10) & 0.16(10) \\
$1558+2642$     &   15.9016(56) &  146.7(3.4) &  -10.4(2.3) &  DA/M&  14560(367)  &   8.74(60) &   1.06(31) &  283(141) &   4 &  320(94)  &   0.32(09) &    38  -    56&    44(17) & 3.45(34) & 0.33(10) & 0.32(24) \\
$1559+0356$$^*$ &    2.2662(12) &   84.3(2.7) &  -50.2(2.0) &  DA/M&  48212(2446) &   7.99(17) &   0.68(09) &  838(106) &  -1$^{**}$ &    -      &   -        &        -      &      -    & -        & -        & -        \\
$1608+0851$     &     9.936(25) &  172.0(4.2) &  -17.6(2.4) &  DA/M&   9844(130)  &   8.60(15) &   0.98(08) &  129(15)  &   6 &  201(93)  &   0.20(08) &    41  -    45&    34(14) & 2.33(11) & 0.20(09) & 0.35(30) \\
$1611+0103$     &    7.2922(56) &   87.2(9.6) &  -11.4(4.7) &  DA/M&  10189(113)  &   7.81(17) &   0.49(10) &  230(23)  &   6 &  192(89)  &   0.20(08) &    23  -    29&    34(15) & 1.65(14) & 0.20(09) & 0.34(33) \\
$1611+4640$     &   1.97678(48) &  273.4(3.9) &    0.2(3.3) &  DA/M&  10307(62)   &   8.04(09) &   0.63(06) &  124(7)   &   5 &  274(140) &   0.25(12) &    51  -    59&   110(52) & 0.49(03) & 0.26(13) & 1.52(50) \\
$1623+6306$     &   53.5624(34) &   74.0(3.2) &   -9.8(2.1) &  DA/M&   9731(138)  &   8.64(16) &   1.00(09) &  179(23)  &   4 &  335(99)  &   0.32(09) &    31  -    34&    23(6)  & 7.89(36) & 0.33(10) & 0.13(08) \\
$1625+6400$     &  5.237716(54) &  100.5(5.2) &  -16.0(3.6) &  DA/M&   8773(76)   &   8.30(15) &   0.79(10) &  155(17)  &   6 &  212(98)  &   0.20(08) &    19  -    22&    24(10) & 1.48(09) & 0.20(09) & 0.49(44) \\
$1646+4223$     &     38.29(21) &  102.9(2.0) &  -14.6(1.4) &  DA/M&  17707(516)  &   7.83(11) &   0.53(06) &  399(27)  &   5 &  300(153) &   0.25(12) &    58  -    75&    49(23) & 5.32(41) & 0.26(13) & 0.13(13) \\
$1705+2109$     &    19.567(61) &   94.8(3.2) &   -0.6(1.8) &  DA/M&  23886(508)  &   7.78(07) &   0.52(04) &  382(19)  &   5 &  276(141) &   0.25(12) &    40  -    44&    46(22) & 3.34(22) & 0.26(13) & 0.20(20) \\
$1718+6101$     &    16.155(44) &   82.8(3.1) &  -23.7(2.4) &  DA/M&  18120(519)  &   7.83(11) &   0.53(06) &  338(24)  &   4 &  464(137) &   0.32(09) &    32  -    38&    49(15) & 3.07(18) & 0.33(10) & 0.25(15) \\
$1731+6233$     &    6.4326(59) &  152.6(4.9) &    0.5(3.0) &  DA/M&  15601(488)  &   7.38(12) &   0.34(04) &  422(37)  &   4 &  479(141) &   0.32(09) &    73  -    90&   143(43) & 1.47(10) & 0.33(10) & 0.42(25) \\
$1844+4120$     &    5.4166(10) &  147.2(3.1) &  -77.7(2.0) &  DA/M&   7554(6)    &   7.49(05) &   0.34(02) &   73(2)   &   6 &   58(27)  &   0.20(08) &    52  -    57&    84(35) & 1.21(08) & 0.20(09) & 0.39(33) \\
$2045-0509$    &  23.53152(84) &   31.1(2.7) &  -16.2(2.1) &  DA/M&  26494(1193) &   7.69(16) &   0.49(07) &  776(83)  &   3 &  855(168) &   0.38(07) &    14  -    17&    24(5)  & 3.94(22) & 0.39(08) & 0.21(09) \\
$2112+1014$     &    2.2152( 1) &    -        &      -      &  DA/M&  19868(489)  &   8.73(10) &   1.06(05) &  206(17)  &   6 &  199(92)  &   0.20(08) &        -      &      -    & 0.87(03) & 0.20(09) & 0.97(79) \\
$2114-0103$    &     9.855(14) &  136.3(2.1) &  -36.6(1.7) &  DA/M&  28064(715)  &   8.11(12) &   0.70(07) &  547(47)  &   3 &  591(116) &   0.38(07) &    42  -    49&    73(15) & 2.26(10) & 0.39(08) & 0.43(17) \\
$2120-0058$    &   10.7754(26) &   57.7(2.9) &  -31.4(1.9) &  DA/M&  16336(291)  &   8.04(06) &   0.64(04) &  218(9)   &   4 &  316(93)  &   0.32(09) &    17  -    19&    28(8)  & 2.33(10) & 0.33(10) & 0.36(20) \\
$2123+0024$     &    3.5835(74) &  149.3(6.2) &   18.3(4.6) &  DA/M&  13432(928)  &   7.31(21) &   0.31(07) &  561(79)  &   6 &  332(153) &   0.20(08) &    42  -    61&    94(43) & 0.85(08) & 0.20(09) & 0.53(52) \\
$2132+0031$     &    5.3336(26) &  181.4(3.6) &   39.0(2.6) &  DA/M&  16336(303)  &   7.53(08) &   0.39(03) &  343(18)  &   4 &  530(156) &   0.32(09) &    $\sim$90   &   148(43) & 1.33(07) & 0.33(10) & 0.50(28) \\
$2149-0717$    &    15.453(11) &  100.7(0.9) &   -1.1(0.6) &  DA/M&  14393(113)  &   8.11(03) &   0.68(02) &  144(3)   &   2 &  220(52)  &   0.44(11) &    39  -    41&    64(1)  & 3.21(02) & 0.44(10) & 0.32(01) \\
$2156-0002$    &  12.07015(25) &   94.7(2.6) &  -39.2(1.5) & DA:/M&           -  &         -  &         -  &    -      &   6 &  177( 82) &   0.20(08) &        -      &      -    & -        & 0.20(99) & -        \\
$2208+1221$     &     45.67(15) &   69.5(1.4) &  -25.7(1.1) &  DA/K&  99575(5434) &   9.02(34) &   1.24(13) &  383(152) &  -1$^{**}$ &    -      &   -        &        -      &      -    & -        & -        & -        \\
$2216+0102$     &    5.0487(51) &  193.6(4.0) &  -11.3(2.8) &  DA/M&  12536(1541) &   7.62(28) &   0.41(14) &  229(37)  &   5 &  360(184) &   0.25(12) &    58  -      &   120(70) & 1.30(18) & 0.26(13) & 0.46(59) \\
$2240-0935$    &    6.2541(27) &  160.7(1.9) &  -15.7(1.4) &  DA/M&  12536(1036) &   7.62(14) &   0.41(06) &  210(18)  &   5 &  231(118) &   0.25(12) &    55  -    83&    99(49) & 1.47(14) & 0.26(13) & 0.41(44) \\
$2243+3122$     &    2.8703(59) &  184.6(5.6) &  -19.4(4.3) &  DC/M&          -   &         -  &         -  &    -      &   5 &  171(87)  &   0.25(12) &        -      &      -    & -        & 0.26(13) & -        \\
$2311+2202$     &   13.9224(17) &  114.8(2.7) &   -1.3(2.0) &  DA/M&  10189(110)  &   9.06(09) &   1.23(05) &  115(12)  &   3 &  474(93)  &   0.38(07) &    29  -    30&    35(6)  & 3.30(08) & 0.39(08) & 0.39(14) \\
$2318-0935$    &   60.8094(27) &   98.2(3.1) &    5.0(1.9) &  DA/M&  22550(603)  &   7.75(09) &   0.50(05) &  460(29)  &   3 &  791(156) &   0.38(07) &   $\sim$90    &    74(15) & 7.49(33) & 0.39(08) & 0.11(04) \\
\noalign{\smallskip}\hline
\end{tabular}
\end{center}
\textbf{Notes.} Radial velocities are measured from the \Na\, doublet, unless marked with $^*$ in which case it comes from the \Ha\, emission line. $\mathrm{K}_{wd}$ are the calculated values in Sect.~\ref{sec:incl}. Errors in the last digits are given in parenthesis. $^{**}$ No spectral type could be derived from fitting the spectrum to template spectra. 
\end{table}
\end{landscape}
\normalsize

%% file: table6.tex
\addtolength{\tabcolsep}{2pt}
\begin{table}
\begin{center}
\caption[Farihi]{\label{t:farihi} 
SDSS-WDMS binaries in common with 
%WDMS binaries with SDSS spectroscopy
%that were also \mbox{imaged} with \textit{HST}/ACS by
  \cite{farihietal10-1}. }
\fontsize{8}{8}\selectfont
\begin{tabular}{llccccccc}
\hline\hline 
\noalign{\smallskip}
WD-Name    & SDSS-Name               &   Type & \# spec  & $\Delta t$ [d] \\ %Notes      \\
\noalign{\smallskip}
\hline
\noalign{\smallskip}
\multicolumn{4}{c}{Not resolved by \textit{HST}} \\
\noalign{\smallskip}
WD$0303-007$ & SDSS\,J$030607.18-003114.4$       &   pceb & 4 & 2991 \\
WD$0908+226$ & SDSS\,J$091143.09+222748.8$       &   pceb & 4 & 5    \\
WD$1051+516$ & SDSS\,J$105421.97+512254.2$       &   pceb & 3 & 1089 \\
WD$1140+004$ & SDSS\,J$114312.57+000926.5$$^{1}$ &   pceb & 3 & 2551 \\
WD$1339+606$ & SDSS\,J$134100.03+602610.4$       &   wide & 3 & 0.03 \\
WD$1433+538$ & SDSS\,J$143443.24+533521.2$       &   pceb & 3 & 1157 \\
WD$1458+171$ & SDSS\,J$150019.33+165914.4$       &   wide & 2 & 0.01 \\ 
WD$1522+508$ & SDSS\,J$152425.21+504009.7$$^{2}$ &   pceb & 3 & 1007 \\
\noalign{\smallskip}
\multicolumn{4}{c}{Resolved by \textit{HST}} \\                  
\noalign{\smallskip}
WD$0257-005$ & SDSS\,J$030024.57-002342.0$       &   wide & 2 & 10   \\ 
WD$1049+103$ & SDSS\,J$105227.72+100337.6$       &   wide & 3 & 0.03 \\  
WD$1106+316$ & SDSS\,J$110843.03+312356.1$       &   wide & 3 & 0.03 \\
WD$1108+325$ & SDSS\,J$111045.90+321447.2$       &   -    & 0 & 0    \\
WD$1133+489$ & SDSS\,J$113609.59+484318.9$       &   -    & 0 & 0    \\
WD$1218+497$ & SDSS\,J$122105.34+492720.5$       &   wide & 3 & 0.03 \\
WD$1236-004$ & SDSS\,J$123836.34-004042.2$       &   wide & 4 & 3197 \\
WD$1419+576$ & SDSS\,J$142105.31+572457.1$       &   wide & 4 & 0.05 \\
WD$1443+336$ & SDSS\,J$144600.72+332849.9$       &   wide & 3 & 0.02 \\
WD$1622+323$ & SDSS\,J$162449.00+321702.0$       &   wide & 5 & 1438 \\
WD$1833+644$ & SDSS\,J$183329.18+643151.7$       &   wide & 9 & 43   \\
\noalign{\smallskip}\hline 
\end{tabular}   
\end{center} 
\textbf{Notes.}
``Type'' is our classification based on the 
radial velocity information. We list the number of spectra used for the 
classification and the time span between them. Two of the systems unresolved by 
\textit{HST} only have SDSS subspectra taken within $<1$\,h.
$^{1}$$\Porb=9.274$ h.
$^{2}$$\Porb=14.15$ h.
\end{table}
\normalsize